# Accessing Electrode Edges: Upright Mounting for Site-Specific Cryogenic Atom Probe Tomography of Lithium-Ion Batteries

Syeda Ramin Jannat[1,2], James O. Douglas[1], Neil Mulcahy[1], Lukas Worch[1], Aigerim Omirkhan[1,2], Zhenyu Guo[2,3], Yuanzhu Zhao[3], Maria-Magdalena Titirici[2,3,4], Ifan E. L. Stephens[1,2], Mary P. Ryan[1,2], Baptiste Gault*[1,5,6], Michele Shelly Conroy*[1]

[1] Department of Materials, Imperial College London, London, SW7 2AZ, United Kingdom
[2] The Faraday Institution, Quad One, Becquerel Avenue, Harwell Campus, Didcot, OX11 0RA, United Kingdom
[3] Department of Chemical Engineering, Imperial College London, London, SW7 2AZ, United Kingdom
[4] Advanced Institute for Materials Research (WPI-AIMR), Tohoku University, 2-1-1 Katahira, Aoba-ku, Sendai 980-8577, Japan
[5] Max Planck Institute for Sustainable Materials, Max-Planck Straße 1, 40237, Düsseldorf, Germany
[6] Université de Rouen Normandie, CNRS, INSA Rouen Normandie, Groupe de Physique des Matériaux, UMR 6634, 76000 Rouen, France

* mconroy@imperial.ac.uk, baptiste.gault1@univ-rouen.fr

## ABSTRACT

Cryogenic atom probe tomography (cryo-APT) is increasingly being used to map the three-dimensional distribution of highly reactive species in heterogeneous materials, including those found in battery systems. However, site-specific preparation of localised features and liquid-solid interfaces remains challenging, particularly when regions of interest (ROIs) are concentrated along electrode edges that are difficult to access using conventional mounting geometries. Here, we present a cryo-APT workflow incorporating an upright-mounting procedure that enables targeted specimen preparation of localised regions using plasma focused ion beam (PFIB) microscopy and milling under cryogenic conditions. We first demonstrate the approach using an electrolyte-wetted graphite electrode, successfully preparing and analysing a specimen containing both an edge-localised electrolyte deposit and the underlying graphite and Cu current collector. We then apply the workflow to a cycled graphite electrode, targeting Li dendrites concentrated around the electrode perimeter and successfully preparing and analysing the targeted Li-rich region. Beyond battery systems, this approach offers potential for site-specific cryo-APT of edge-localised features across a broader range of material systems, where conventional specimen preparation geometries may limit access to ROIs.

# 1 INTRODUCTION

Atom probe tomography (APT) offers three-dimensional chemical analysis at nanoscale spatial resolution and has increasingly been applied to the characterisation of complex interfaces in lithium-ion batteries (LIBs)[1–4]. However, the preparation and analysis of battery materials by APT remain challenging[2]. Electrode components and their interfaces can be highly sensitive to air, moisture and ion-beam exposure[5–7], while the porous and heterogeneous nature of composite electrodes introduces further difficulties during focused ion beam (FIB) specimen preparation and subsequent field evaporation during the APT analysis[8,9]. Cryogenic workflows have therefore become increasingly important for preserving sensitive materials and their interfaces throughout specimen preparation, transfer and analysis[10,11].

Recent developments in cryogenic and ultra-high vacuum (UHV) preparation and transfer have enabled sensitive specimens to be maintained under controlled conditions between preparation and APT analysis. Integrated cryogenic transfer systems have been developed to minimise atmospheric exposure during specimen handling[12], while cryogenic FIB techniques have enabled the site-specific preparation of frozen liquid-solid interfaces for APT[10,13,14]. Much of the development of cryogenic FIB preparation has taken place in the field of biological scanning and transmission electron microscopy (SEM, TEM, respectively), where cryogenic milling and lift-out have enabled selected regions of frozen hydrated cells and tissues to be prepared for cryo-TEM[15]. Similar approaches have been developed for the study of sensitive liquid-solid interfaces in materials science. Schreiber *et al.*, for example, demonstrated the site-specific cryogenic preparation of a frozen water-glass interface for APT[16], while Schwarz *et al.* used cryogenic preparation to arrest corrosion at a Mg alloy-water interface, preserving the reacting interface for subsequent correlative cryo-APT and EM[10].

In lithium-based battery systems, cryogenic preparation has facilitated the characterisation of intact electrode–electrolyte interfaces and Li plating using scanning transmission electron microscopy (STEM)[17], with more recent correlative workflows extending these capabilities to electrolyte-containing interfaces for cryogenic APT[4]. For example, Mulcahy *et al.* demonstrated a workflow linking *in situ* liquid-cell TEM with cryogenic plasma FIB preparation and APT, enabling an electrochemically formed liquid-solid interface to be frozen and subsequently analysed in three dimensions[4]. Related approaches have also been used to investigate phase transformation during electrochemical cycling using *in situ* hard X-ray measurements and cryo-EM[18]. These workflows demonstrate the value of retaining electrochemical interfaces for structural and chemical analysis, but preserving the material alone does not ensure that a particular region of interest (ROI) can be successfully prepared or analysed. APT requires a needle-shaped specimen containing the targeted material, with specimen orientation being particularly important for multiphase systems. Consequently, the initial mounting configuration must provide both access to the ROI and a suitable orientation for subsequent APT analysis.

Conventionally, sections of planar battery electrodes can be mounted flat beneath the spring-loaded clip of a cryogenic sample holder. This configuration provides access to the principal electrode surface and is suitable for targeting features distributed across this face. However, its application becomes more limited when the ROI is located at the electrode perimeter, where localised deposits or buried interfaces may be present. An example of this includes edge-localised Li dendrites, where spatially heterogeneous Li plating is concentrated at or near the electrode perimeter rather than distributed across its principal surface. Although such features may be accessible for FIB lift-out from a flat-mounted electrode, the resulting specimen may contain the constituent phases arranged laterally relative to the field-evaporation axis. A sequential arrangement is generally preferable, as it facilitates the successive analysis of different phases and reduces complications arising from their differing evaporation behaviours[19–21]. Differences in evaporation field between adjacent phases can introduce local magnification effects and ion trajectory aberrations, potentially complicating the reconstruction of their spatial relationships[1]. Also, the high electrostatic stresses experienced by specimens during APT data acquisitions can lead to premature fractures, particularly where mechanically weak regions or interfaces are present[22]. Alternative specimen orientations have been employed in FIB preparation, particularly for TEM, using specialised mounting arrangements[23,24]. However, these may not provide sufficient mechanical support for planar electrodes or be readily compatible with cryogenic transfer hardware. An alternative mounting approach is therefore needed that provides access to edge-localised features while enabling the preparation of appropriately oriented APT specimens.

Here, we present an upright-mounting procedure for the site-specific cryogenic FIB preparation of LIB electrodes enabling subsequent APT analysis. The method uses folded Cu tape to secure the electrode perpendicular to the cryogenic specimen holder, exposing its edge for targeted FIB lift-out while providing mechanical stability during cryogenic handling and transfer. By using readily available laboratory materials and existing specimen holders, this approach also avoids the need for specialised mounting equipment or modified FIB stages, making it straightforward to implement within established cryogenic workflows. This configuration enables edge-localised features and their underlying material layers to be incorporated into APT specimens with the constituent phases aligned along the field-evaporation axis.

The procedure is first demonstrated using a graphite electrode wetted with uncycled electrolyte as a model system, followed by its application to an over-lithiated graphite electrode containing edge-localised Li dendrites. The approach provides an accessible and adaptable extension to existing cryogenic APT workflows, with wider applicability to other sensitive materials containing edge-localised or buried interfaces.

# 2 MATERIALS AND INSTRUMENTATION

## 2.1 Sample Preparation and Electrochemistry

To prepare the graphite electrodes, a slurry containing 5-µm graphite powder (99%, Graphene Supermarket, Ronkonkoma, NY, USA), Super P carbon black (>99%, Imerys Graphite & Carbon, Bodio, Switzerland), and polyvinylidene fluoride (PVDF, Merck KGaA, Darmstadt, Germany) in a mass ratio of 90:2:8 was first formulated inside a fume hood. PVDF was used as a 5 wt.% solution in N-methyl-2-pyrrolidone (NMP, anhydrous 99.5%) as the solvent by mass, supplied by Merck KGaA (Darmstadt, Germany). The components were initially mixed manually to form a coarse mixture, followed by the use of a THINKY ARE-250 planetary centrifugal mixer (THINKY Corporation, Tokyo, Japan) to obtain a homogeneous slurry. The resulting slurry was cast onto 12 µm-thick Cu foil (Goodfellow Cambridge Ltd., Huntingdon, UK) using a spatula and adjustable wet-film applicator, and the electrodes were then dried overnight at 353 K inside a fume hood, followed by vacuum drying at 353 K for 2-3 hours before use or transfer to an Ar-filled glovebox for storage. For the matchstick electrodes used in this study, the Cu current collector strips were cut to dimensions of 10 mm x 1 mm (L x W), with a 1 mm x 1 mm square of active material remaining at the top (0.01 $cm^2$ geometric electrode area). The graphite anodes prepared using this method had a typical mass loading of ~7.7 mg $cm^{-2}$.

For full-cell configurations, a commercial $LiNi_{0.8}Mn_{0.1}Co_{0.1}O_2$ (NMC811) electrode (Li-Fun Technology Corporation Limited, Zhuzhou, China) was used as the cathode, with a mass loading of 15.5 mg $cm^{-2}$, and cut to the same dimensions as the graphite electrode. Where Li discs (PI-KEM Ltd., Tamworth, UK) were used, these had a thickness of 0.25 mm and diameter of 4 mm.

To demonstrate the upright-mounting procedure, this was first trialled on a pristine graphite electrode that was wetted with uncycled electrolyte, 1.0 M $LiPF_6$ in 1:1 v/v ethylene carbonate to ethyl methyl carbonate (EC:EMC, Merck KGaA, Darmstadt, Germany) as a model system. The electrode was positioned upright in an Ar-filled glovebox with $H_2O$ < 0.5 ppm and $O_2$ < 0.5 ppm (MBraun, Garching, Germany), and ~3 droplets of electrolyte were pipetted across the active area of the electrode, ensuring that the electrode was fully wetted. Care was taken to saturate the upper edge of the electrode with electrolyte, allowing this region to act as a proxy for an edge-localised feature. The electrode was then left to wet for approximately 1 hour to facilitate electrolyte infiltration into the porous composite electrode structure.

To demonstrate the workflow on a more scientifically relevant application, an *operando* commercial ECC-Opto-10 cell (EL-Cell GmbH, Hamburg, Germany) was used to probe edge-localised Li plating on a graphite electrode[25]. The cell was assembled in a side-by-side configuration, using graphite, NMC811, and Li as the working, counter, and reference electrodes, respectively. After leaving to rest for 3 hours, the cell was connected to a VHX-X1 digital microscope with 20x – 6000x zoom range (Keyence Corporation, Osaka, Japan), and PalmSens4 potentiostat (PalmSens BV, Houten, Netherlands) which had a -5 V – 5 V potential range and 100 pA – 10 mA current range. The details of this cell, its configuration, and material dimensions are given in Supplementary Information (Fig. S1). One formation cycle was performed between 1.0 – 0.005 V vs. $Li^+/Li$ (current density of 0.24 mA $cm^{-2}$, C/10 rate), before a plating sequence was carried out by lithiating the graphite to -0.05 V vs. $Li^+/Li$ using a current density of 0.30 mA $cm^{-2}$. Photographs were taken every 2.5 minutes. The electrochemical experiment was terminated once sufficient edge-localised Li plating was observed, after which the electrode was immediately prepared for cryogenic transfer. All electrochemical measurements were performed at room temperature.

**2.2 Cryogenic Sample Retrieval and Transfer**

For the cycled cell, following completion of the electrochemical experiment, this was then transferred into an Ar-filled glovebox and disassembled. The graphite electrode was carefully extracted and transferred out of the glovebox before being immediately placed onto a pre-cooled copper block ('dual-puck stage', Oxford Atomic Ltd., Oxford, UK) that was partially submerged in liquid nitrogen ($LN_2$)[4]. The electrode was slowly cooled before being transferred into a vented centrifuge tube and fully submerged in $LN_2$ for cryogenic storage. For the uncycled sample, the electrode retrieval step was not required, but the same cooling and cryogenic storage procedure was followed.

For subsequent specimen preparation, the frozen electrode was transferred under $LN_2$ into an $N_2$-filled glovebox (GB-1200E glovebox, Sylatech GmbH, Walzbachtal, Germany), which had typical $O_2$ and $H_2O$ readings of <5 ppm. The equipment includes two antechambers of varying sizes to transfer from the outside environment to inside the glovebox. Inside, there is also an $LN_2$ basin (1.5 L) which can be filled using a pressurised dewar (50 L APOLLO system, Cryotherm GmbH & Co. KG, Kirchen, Germany). The glovebox comes fitted with a load-lock chamber which can be vented and used to move a stage-containing elevator down that holds the mounted puck. Once on the pre-cooled elevator, this can be transported up into the load-lock and to a vacuum cryo-transfer module (VCTM, Ferrovac AG, Zürich, Switzerland) using a 500-mm wobblestick. The pressure in this module is in the range of $10^{-10}$ mbar, maintained using a non-evaporable getter cartridge and ion-pump element, and the carried sample is kept at cryogenic temperatures using an $LN_2$ dewar. This VCTM is compatible with standard APT cryo-pucks (CAMECA Instruments Inc., Madison, WI, USA) and enables the transfer of samples between the plasma focused ion beam (PFIB), glovebox, and the atom probe under UHV, temperature-controlled conditions[12].

**2.3 Cryogenic PFIB Specimen Preparation**

A Helios Hydra CX Plasma FIB (Thermo Fisher Scientific, Hillsboro, OR, USA), retrofitted with an Aquilos cryo-stage, was used for the cryogenic specimen preparation in this study. The instrument was equipped with an Easy-Lift tungsten micromanipulator which was compatible with cryogenic operation, enabling site-specific lift-out of material. Both the stage and manipulator were maintained at 113 K by circulating gaseous nitrogen through a heat exchanger system at a flowrate of 180 mg $s^{-1}$, and an $LN_2$-filled dewar, with cooling from room temperature to 113 K taking approximately 1 hour. Temperature regulation was achieved using a Model 335 cryogenic temperature controller (Lake Shore Cryotronics, Westerville, OH, USA) and the corresponding heaters.

The PFIB is also fitted with a FerroLoader docking station (Ferrovac), allowing the pre-cooled VCTM to be docked directly to the instrument, and the cryogenic sample transferred directly from the VCTM onto the pre-cooled PFIB cryo-stage. For ion beam use, eucentric height was achieved at a 52° stage angle (normal to the FIB) and Xe was used as the primary plasma source (although Ar is often useful for polishing carbon-containing materials)[26,27]. The sample was mounted onto a dual-puck holder for SEM or FIB (Oxford Atomic Ltd., Oxford, UK), which enables two cryo-pucks (CAMECA Instruments Inc., Madison, WI, USA) to be placed onto the stage at once. For this work, one puck holds the frozen graphite sample, and the other a Si 22-microtip coupon (M22, CAMECA Instruments Inc., Madison, WI, USA) used as a support for APT specimens. Details of the specimen preparation procedure, including the ion beam acceleration voltages, milling currents, pattern dimensions, and needle sharpening steps, are described in the Results and Discussion section.

**2.4 APT Measurements**

Following PFIB preparation, specimens were transferred under cryogenic and vacuum conditions to the atom probe using the VCTM and were loaded onto the equipment using the FerroLoader docking station. This enables the transfer of a puck from the VCTM to the analysis chamber of the local electrode atom probe (LEAP) 5000 XR (CAMECA Instruments Inc., Madison, WI, USA) through the use of a 'piggy-back' puck (which, as the name suggests, 'carries' the puck through this process). The puck was maintained under vacuum and at cryogenic temperatures[12,28,29].

APT measurements were performed in laser-pulsing mode at a specimen base temperature of 30 K, using a laser pulse energy of 50-60 pJ, a target detection rate of 0.3% and a pulse repetition rate of 125 kHz. APT dataset reconstruction and analysis (compositional information, concentration profiles, and contour maps) were carried out using IVAS 6.3 within AP Suite 6. Post-processing and plotting of data was conducted using OriginPro.

# 3 RESULTS AND DISCUSSION

## 3.1 Cryogenic Workflow and Development of an Upright-Mounting Procedure

The complete cryogenic workflow used to prepare LIB electrodes for APT is summarised in Fig. 1. The workflow was designed to maintain the electrode under cryogenic conditions and controlled-atmosphere from retrieval through to APT analysis, thereby limiting modification of the reactive and air-sensitive sample during specimen preparation and transfer. For cycled samples, following cycling to the desired electrochemical state of charge, the cells were disassembled and the graphite electrode was retrieved (Fig. 1(a)). For the uncycled proof-of-concept sample, the graphite electrode was instead wetted directly with electrolyte prior to freezing (more details in Section 2.1).

The electrode was then immediately transferred out of the glovebox onto a copper block which had been pre-cooled with $LN_2$ (boiling point = 77.4 K) for several minutes beforehand until bubbles subsided, ensuring that the block was sufficiently cold for freezing. The concept of using cold blocks for freezing samples has been well-reported in the literature, with copper and silver commonly used due to their high thermal conductivities (400 W $m^{-1}$ $K^{-1}$ and 429 W $m^{-1}$ $K^{-1}$, respectively)[30,31]. The electrode was kept on the cold block until the electrolyte solidified (~15 – 30 seconds, appearing fully matte upon freezing). Ideally, this freezing process would also take place under inert conditions as LIB systems, including the electrolyte and solid-electrolyte interphase (SEI) species, the passivating layer formed at the electrode surface through electrolyte decomposition, are air-sensitive[32]. Reports on graphitic electrodes, however, have shown minimal differences between the SEI constituents present in air-free environments vs. when exposed to air for <2 mins. It has also been shown that SEI degradation with air contact is progressive with time and hence, artefacts are expected to be minimal under the short exposure times, particularly because the residual electrolyte layer on the electrode is likely to prevent direct contact between the delicate SEI components and atmosphere[4,6,33]. A more detailed step-by-step workflow for the electrode extraction post-cycling and freezing process is provided in Fig. S2 in Supplementary Information.

Once frozen, the electrode was picked up, placed into a vented centrifuge tube, and stored under $LN_2$, as shown in Fig. 1(b). Prior to further preparation, the $LN_2$ basin in the glovebox was filled (see Fig. S3) and the frozen electrode was mounted onto a cryo-puck and positioned onto the elevator, illustrated in Fig. 1(c) – (e), which had been pre-cooled to ~113 K using the pressurised dewar and moved down. The mounted sample was subsequently transferred into the VCTM, maintained at ~93 K (Fig. 1(f)). Photographs are provided in Fig. S4 in Supplementary Information. The VCTM was then detached from the glovebox and attached to the PFIB (Fig. 1(g)) via the FerroLoader docking station. Fig. 1(h) shows the site-specific specimen preparation that could be performed using this equipment under cryogenic conditions. Following preparation, the specimen was transferred to the LEAP for APT data acquisition (Fig. 1(i)).

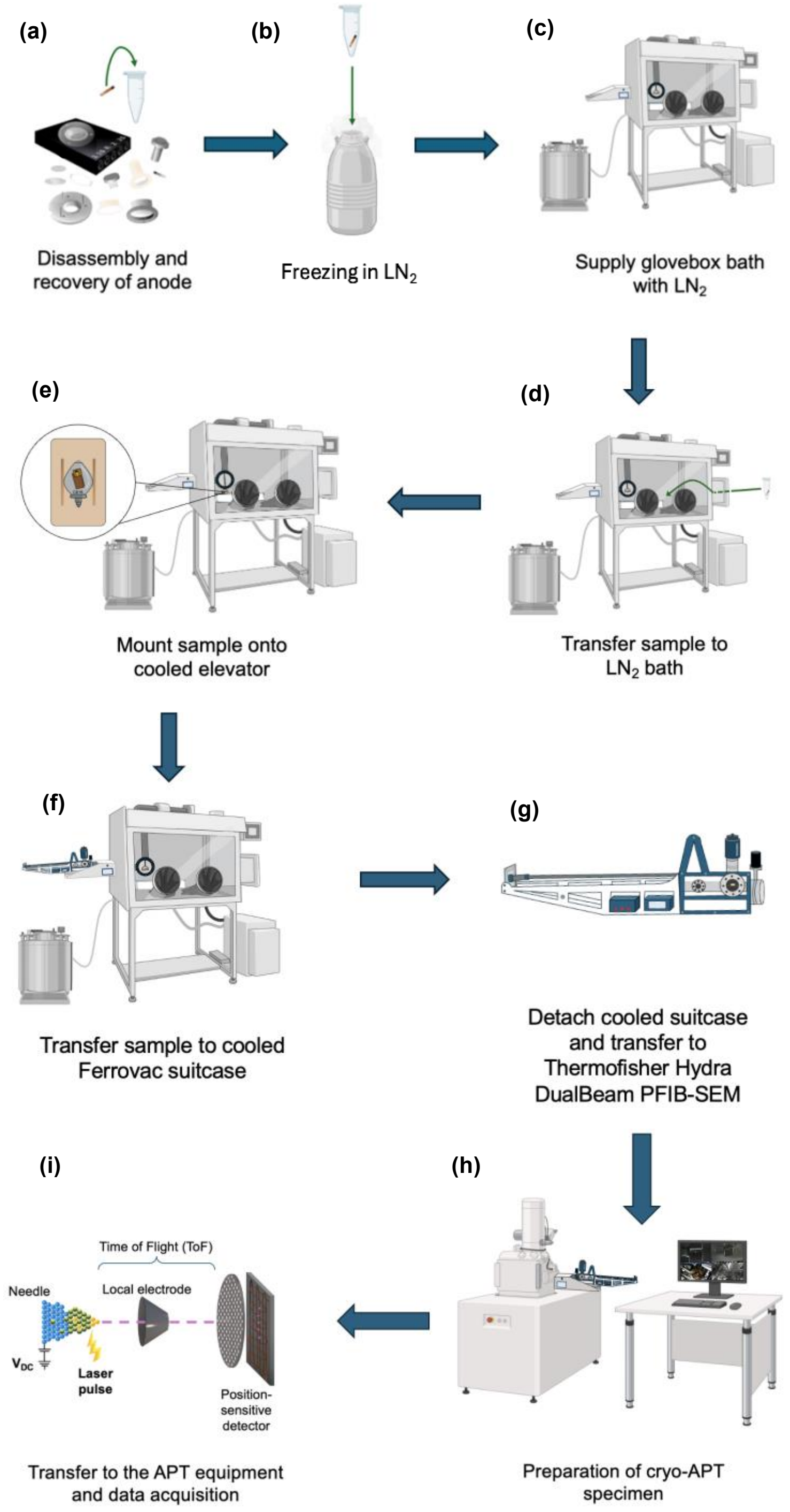

**Figure 1 – Overview of the cryogenic workflow.** (a) Electrode retrieval following cell disassembly, (b) freezing on an $LN_2$-cooled block and storage in $LN_2$, (c) filling the glovebox basin with $LN_2$, (d) transfer of the frozen sample into the $LN_2$ bath, (e) flat or upright mounting onto a cryo-puck, (f) transfer into the pre-cooled VCTM, (g) transfer to the PFIB, (h) cryogenic preparation of the APT needle and (i) transfer of the VCTM (with APT needle) to the LEAP 5000 XR equipment and subsequent data acquisition. Image partially created in BioRender. Jannat, S.R. (2026) https://BioRender.com/clwromc.

Within this workflow, the specimen mounting step in Fig. 1(e) is particularly important because the orientation of the bulk electrode determines the accessibility of specific regions during site-specific PFIB preparation. Conventional mounting positions a section of the planar electrode flat beneath the spring-loaded clip of the cryo-puck. This configuration provides direct access to the principal electrode face and is suitable where the ROI lies across the surface. In contrast, features confined to the electrode perimeter are less readily accessible from this geometry. An alternative orientation was therefore required to expose the electrode edge directly to the ion beam.

Fig. 2 compares the conventional flat-mounted geometry with the upright configuration developed here. Fig. 2(a) shows an optical image of a cycled graphite electrode with the Li dendrites concentrated around its perimeter, illustrating an example of the edge-localised features motivating the upright-mounting approach in this study. The *operando* set-up used to prepare this electrode is shown in Fig. S1. In the flat-mounted arrangement, the electrode is secured under the cryo-clip with its broad face exposed, allowing FIB access to ROIs across the principal electrode surface, as shown in Fig. 2(b). In contrast, Fig. 2(c) shows the upright configuration, in which the electrode is oriented perpendicular to the cryo-puck surface.

The step-by-step mounting method is then demonstrated using a separate graphite electrode without Li dendrites, illustrated in Fig. 2(d) – (h). First, Cu tape was placed beneath the electrode, as shown in Fig. 2(d), before the sides were folded around the sample to secure it (Fig. 2(e)). Fig. 2(f) then shows this arrangement positioned flat underneath the cryo-puck clip. Next, the Cu tape was manipulated to orient the electrode perpendicular to the cryo-puck, with a front view of the resulting geometry shown in Fig. 2(g). In this configuration, the Cu tape serves several purposes: it acts as a mechanical brace for the electrode itself whilst also conferring additional thermal conductivity to the sample which is a useful option for insulating materials[34,35]. It also remains ductile at cryogenic temperatures so the sample can be manipulated freely inside the $LN_2$-containing glovebox. Alternative methods for mounting samples vertically in the FIB have previously been reported, particularly for TEM applications[5,24,36]. However, these arrangements are generally designed to accommodate specific specimen dimensions, and may not provide sufficient mechanical support for securing planar electrode sections in an upright configuration, as required in this study. As such, the procedure developed here bypasses the reliance on specialised, geometry-specific, (often) non-adaptable equipment, providing a simple, low-cost, and accessible means of securing electrodes upright while maintaining compatibility with the VCTM, FIB, and APT instrumentation.

To test the mechanical stability of the mounted sample, it was transferred several times between the different instruments of the workflow. It was found that the electrode remained fixed at approximately 90$^{o}$ to the cryo-puck following repeated transfers, remaining visible in this orientation in the LEAP analysis chamber, shown in Fig. 2(h). This demonstrates that, for this Cu tape and mounting configuration used here, sufficient mechanical support was provided to maintain the electrode in its upright position throughout the cryogenic workflow without the need for a dedicated specimen holder. Importantly, orienting the electrode in this way provides direct access to the electrode edge, but also alters its position relative to the PFIB columns compared with conventional flat mounting. Additionally, the cooling requirements of cryogenic manipulators may restrict their rotational capabilities, making it more challenging to adjust the orientation of extracted specimens following lift-out. A corresponding specimen preparation procedure was therefore developed to enable site-specific isolation and lift-out of regions from the exposed electrode edge.

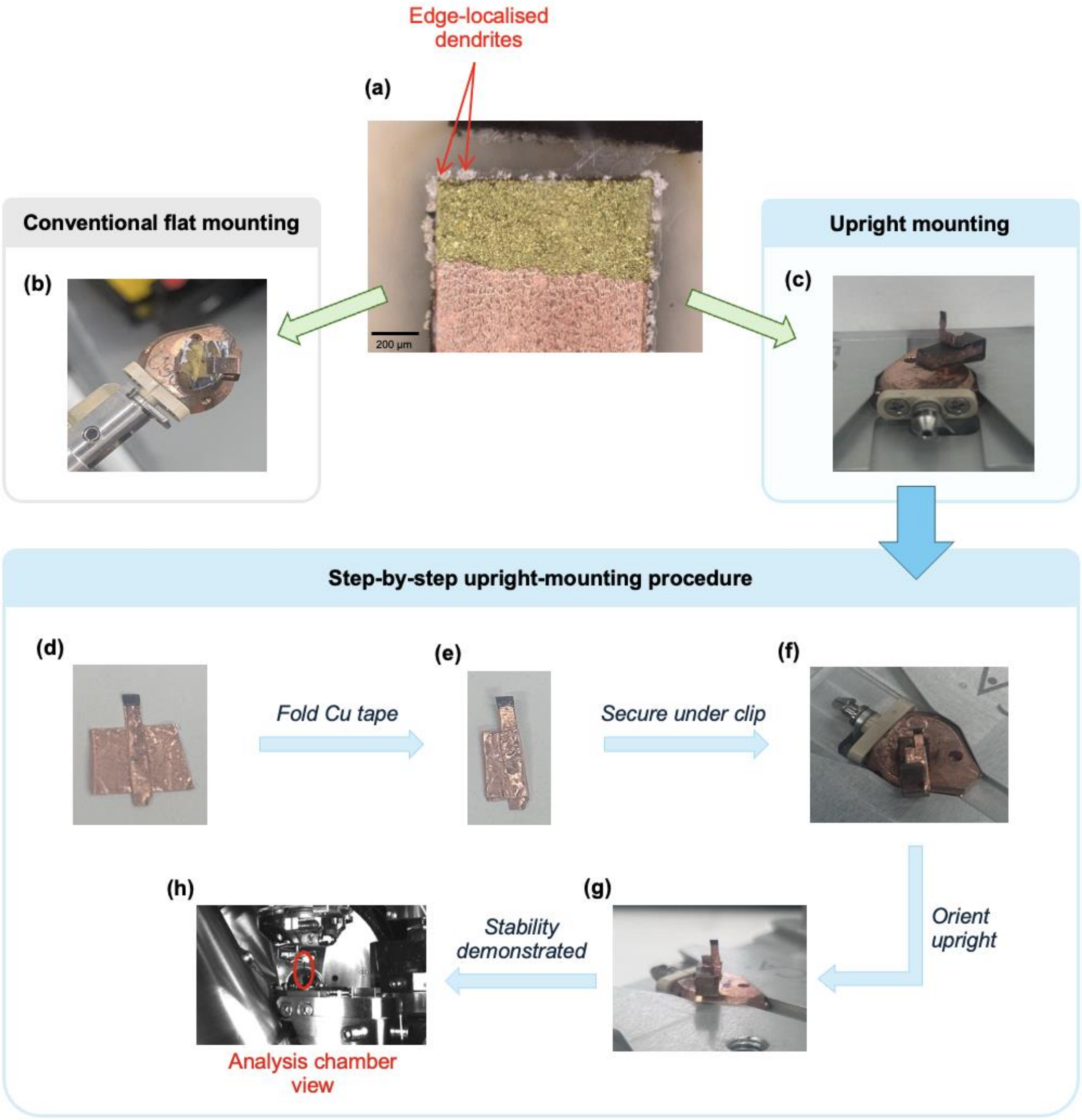

**Figure 2 – Flat- and upright-mounting configurations for cryogenic specimen preparation.** (a) Optical image of a cycled graphite electrode showing Li dendrites concentrated around the electrode perimeter, illustrating an example of edge-localised features targeted by the upright-mounting approach. Mounting configurations: (b) graphite electrode mounted flat beneath the spring-loaded clip of a cryo-puck, and (c) graphite electrode mounted upright, perpendicular to the cryo-puck surface. Step-by-step upright-mounting procedure: (d) Cu-tape positioned beneath the electrode, (e) Cu tape folded around the electrode to secure it, (f) electrode and Cu tape positioned flat beneath the spring-loaded clip of the cryo-puck, (g) electrode oriented upright, perpendicular to the cryo-puck surface, shown from the front, and (h) upright-mounted electrode inside the LEAP analysis chamber following transfer, with its position highlighted by the red circle. The electrode remained fixed upright at 90$^{o}$ to the cryo-puck following repeated transfers between equipment, demonstrating the mechanical stability and robustness of the mounting configuration.

This sequence is summarised in Fig. 3 for a cycled graphite example with Li dendrites along the edge of the electrode. The electrode was initially positioned upright at 0° stage tilt (Fig. 3(a)), such that the mounted sample was parallel to the SEM column. The stage was then rotated by +90° such that the Cu current-collector side was exposed to the ion beam (Fig. 3(b)). The next step involved identifying an ROI that was relatively flat (battery electrodes can often be heterogeneous in morphology, especially graphite) and, ideally, dense and well-packed, as porosity in APT specimens can present substantial challenges. Open voids compromise the mechanical stability of needles, increasing the likelihood of tip bending during preparation in the FIB or fracturing prematurely in APT. Pores are also known to result in artefacts in the trajectories of the field-evaporated ions, local magnification effects, and unreliable quantification[37,38]. Electrodes were wetted with electrolyte prior to cycling, reducing the extent of electrolyte-accessible void space remaining unfilled within the composite graphite electrode[8,9].

Prior to cooling the stage, a Cr block (~20 μm x 5 μm x 10 μm) was attached to the manipulator under room-temperature conditions. This was subsequently used to sputter-deposit a thin metallic coating over the selected ROI on the electrode surface, providing protection for the sensitive electrolyte or Li-containing components against ion beam damage[39,40]. Conventionally, protective coatings are deposited using a gas injection system (GIS), which typically employs organometallic precursors containing metals such as Pt or W[41–43]. Under cryogenic conditions, however, these precursors can condense uncontrollably across the sample surface, resulting in non-localised deposition[4]. To achieve localised Cr deposition, the stage was maintained at 0° stage tilt, and the manipulator was positioned such that the Cr block was ~10 – 20 μm away from the ROI surface. The Cr block was then sputtered using sequential cross-section milling patterns at 30 kV and 1.0 nA, with a single pass. The resulting Cr-coated region could be distinguished from the surrounding uncoated surface by a clear change in image contrast, as shown in Fig. S5.

Next, the stage was tilted to -7° to provide suitable ion-beam access to the ROI, allowing a 'finger' to be prepared, as shown in Fig. 3(c), at 30 kV and 60 nA (pattern dimensions of 30 μm x 60 μm x 50 μm). Preparation of this finger was necessary to isolate the ROI (e.g., an edge-localised Li dendrite) so that it could be accessed more easily and lifted out using the manipulator. A standard lift-out was then initiated at 52°, first by preparing a cantilever at 30 kV and 15 nA (dimensions: 30 x 10 x 10 $\mu m^3$) ion beam conditions (Fig. 3(d)). Graphite is a significantly challenging material to work with due to the strong $sp^2$ covalent bonds present, as well as the highly anisotropic structure, with basal and edge planes yielding different sputtering rates[44,45]. Hence, multiple passes were often required, which were carried out at lower currents (≤ 4.0 nA) to minimise cumulative beam damage. An undercut was then prepared at 0° (30 kV, 4.0 nA) and the manipulator was positioned to contact the sample. At the tip of the manipulator, a portion of Cr was retained as this was found to yield a stronger adhesion to the graphite than the W manipulator. Once attached, redeposition sputtering (30 kV, 30 pA, cross-section milling, single pass) was used to secure this Cr-sample connection, consistent with previously reported redeposition-based attachment procedures, including those outlined by Woods *et al.* and Douglas *et al.*[39,40,46]. A final cut was then performed at 30 kV, 1.0 nA to free the cantilever from the bulk electrode, as shown in Fig. 3(e).

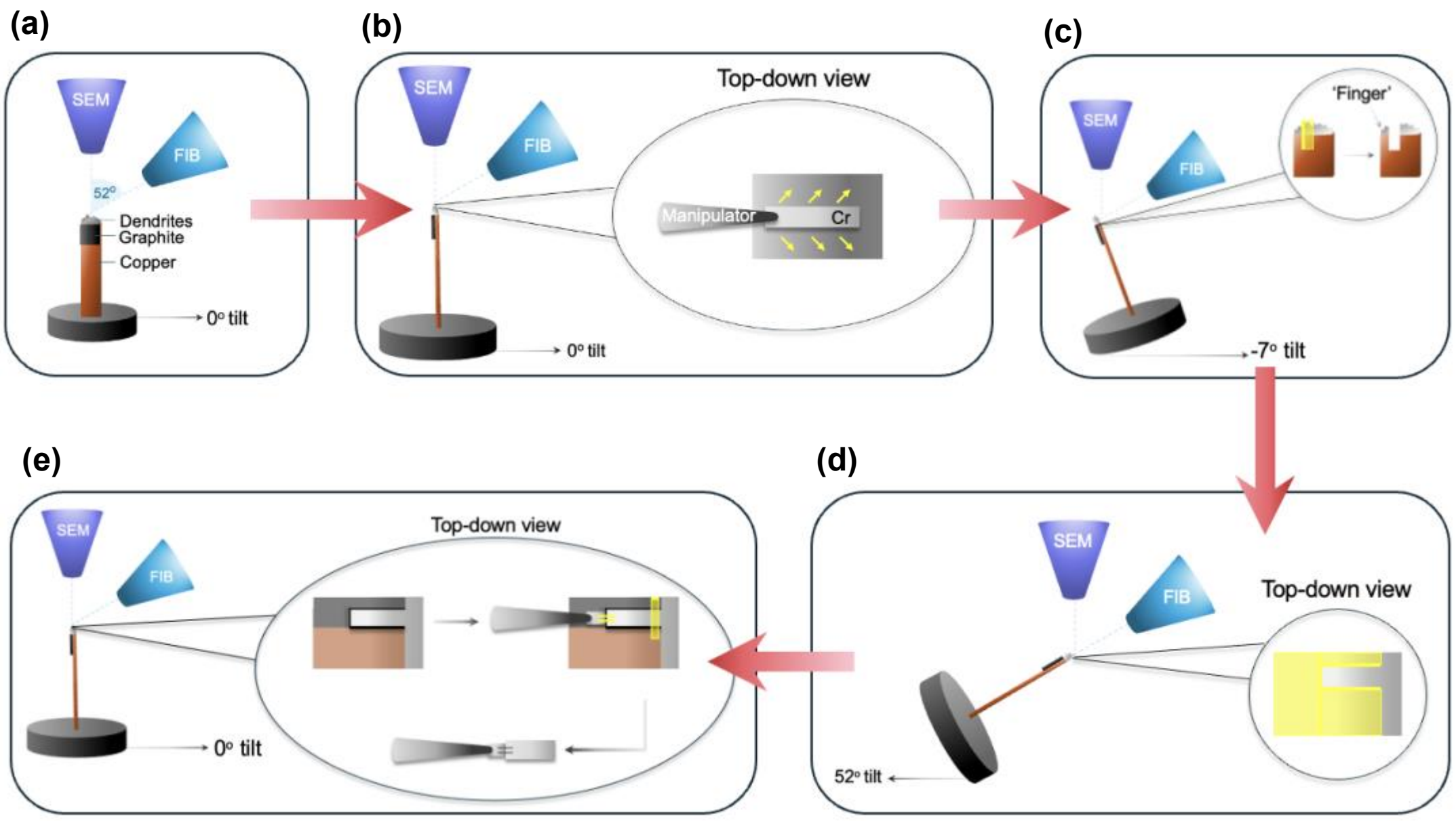

**Figure 3 – Site-specific preparation and lift-out of an ROI from an upright-mounted electrode.** (a) Upright-mounted electrode at 0° stage tilt with the graphite, Cu current collector and edge-localised feature (as an example here) clearly labelled, (b) stage rotated by +90° to expose the Cu current collector to the ion beam, followed by Cr deposition over the ROI, (c) stage tilted to -7° for isolation of the ROI as a protruding 'finger', (d) preparation of the cantilever at 52° stage tilt, and (e) undercutting at 0° followed by attachment of the Cr-tipped manipulator and lift-out.

Fig. 4 shows the next part of the APT specimen preparation process. The lifted-out cantilever was first transferred to a pre-prepared Si post (see Fig. 4(a)). This post was pre-prepared at room-temperature conditions by milling the top part of the post at 0° tilt and 30 kV, 4.0 nA ion beam conditions, while leaving a portion of the post remaining on the side to improve the contact area with the cantilever, thus acting as a support (more information on this is provided by Mulcahy *et al.*[47]).

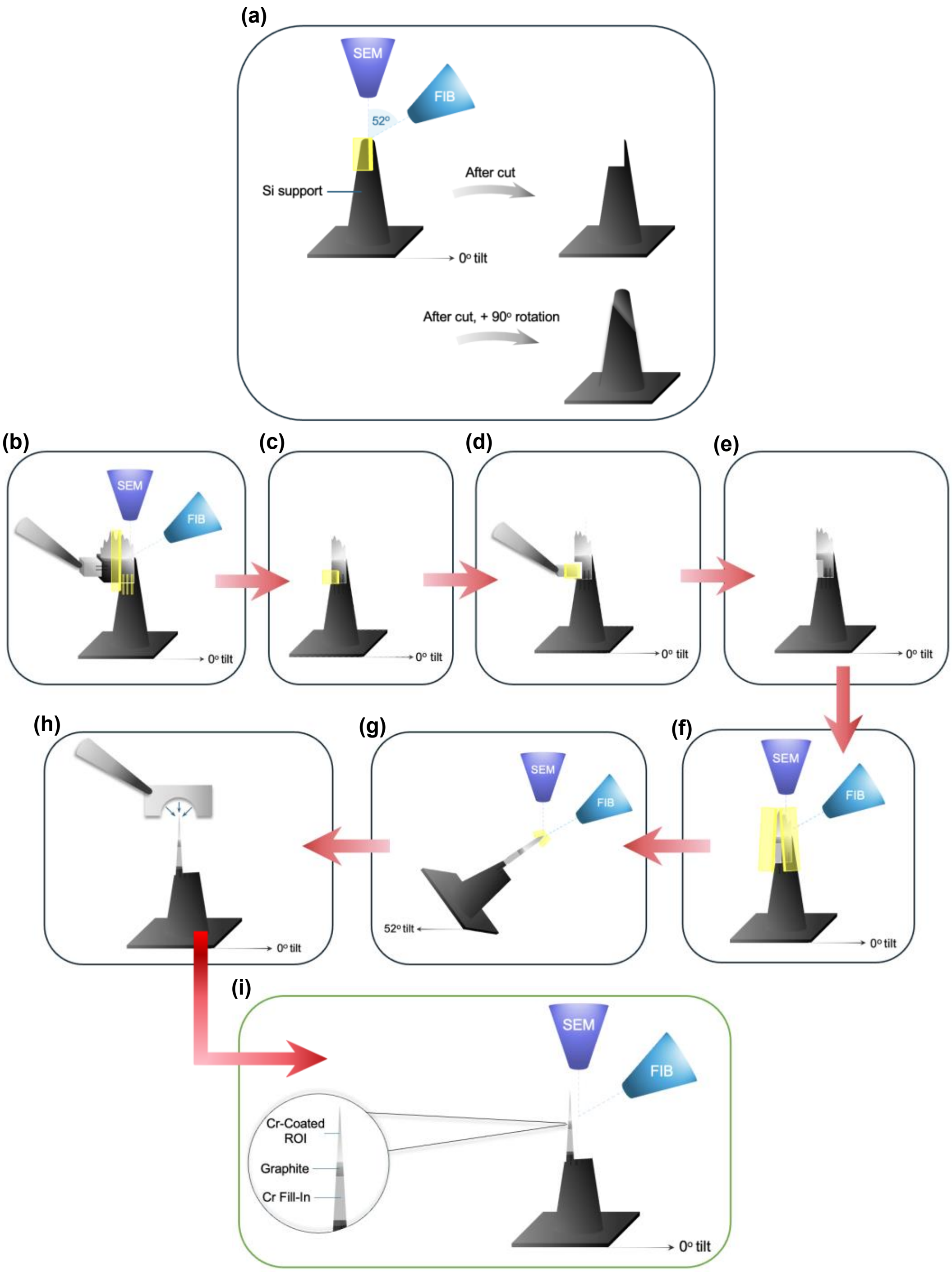

**Figure 4 – Preparation of the Si post and subsequent preparation of the APT specimen following site-specific lift-out.** (a) Preparation of the Si post at room temperature by removing a portion of the post while retaining a side support to improve specimen stability during attachment, (b) lifted-out specimen positioned against the pre-prepared Si post and secured by redeposition sputtering, (c) preparation of a side opening for Cr fill-in, (d)-(e) Cr fill-in procedure to secure the specimen to the post, (f) removal of excess material at 0° to produce a specimen with an approximate tip size of 1 µm x 1 µm, (g) stage tilted to 52° for successive milling to sharpen the specimen, (h) Cr coating of the prepared needle, and (i) final Cr-coated APT specimen, with the inset showing the Cr-coated region, graphite, and Cr fill-in.

The sample was subsequently attached to the Si post (Fig. 4(b)), again using redeposition sputtering on all sides, before a side hole was cut out (Fig. 4(c), using the process outlined by Douglas *et al.*[39]), and Cr was used to fill this gap in, as shown in Fig. 4(d) – (e). The purpose of this fill-in step was to secure the sample to the post as the redeposition sputtering acted only as temporary attachments. The use of Cr also improves the mechanical stability of the specimen and confers additional conductivity[2,48]. The fill-in process was conducted at 30 kV, 30 pA – 0.3 nA ion beam conditions.

Once the gap had been sufficiently filled, the specimen was shaped into an initial APT needle by removing excess material using rectangular milling patterns at 0° stage tilt, 30 kV, 4.0 nA, as shown in Fig. 4(f). The specimen was first reduced to ~1 μm x 1 μm before tilting the stage to 52° and continuing with sequential milling patterns to achieve a tip diameter below 300 nm, ideally below 100 nm (Fig. 4(g))[1]. The ion beam current was progressively reduced from 4.0 nA to 30 pA as the needle diameter decreased to minimise beam-induced damage[49]. Unlike conventional APT tip preparation, which typically uses annular milling, rectangular patterns were used here to allow greater control over the material removal by milling each side individually.

Finally, Cr was sputter-coated onto all four sides of the finished needle for 20 s per side at 30 kV and 30 pA, shown in Fig. 4(h), with the inner edge of the Cr semi-circle positioned ~1.5 μm from the needle apex. This coating is an important step as it improved the electrical conductivity of the needle and helps to prevent *in situ* lithium migration, a common artefact reported to occur during APT data runs of Li-containing materials. It also reduces thermal and electrostatic stresses on the needle, which could otherwise lead to premature specimen fracture[50,51]. A schematic of the example material system outlined here is shown in Fig. 4(i).

The feasibility of the complete preparation procedure was then assessed using an uncycled electrolyte-wetted graphite electrode as a model system.

### 3.2 Validation Using an Electrolyte-Wetted Graphite Test System

The upright-mounting procedure was first demonstrated on a pristine graphite electrode deliberately wetted with uncycled electrolyte (Fig. 5). Fig. 5(a) – (g) shows the isolation of a site-specific ROI from the electrode edge, followed by finger preparation, lift-out, and attachment to the Si post, while Fig. 5(h) – (k) illustrates the subsequent Cr fill-in and successive sharpening steps. Importantly, the edge-localised ROI could be retained throughout the preparation process without reorienting the bulk electrode into a conventional flat-mounted configuration. This is especially important for multiphase regions at the electrode edge, as it allows the different phases to be retained through the depth of the specimen rather than distributed laterally across the lift-out. The final Cr-coated specimen had a tip diameter of ~250 nm, as shown in Fig. 5(l), demonstrating that the upright geometry could be used to prepare an APT-compatible specimen directly from the electrode perimeter.

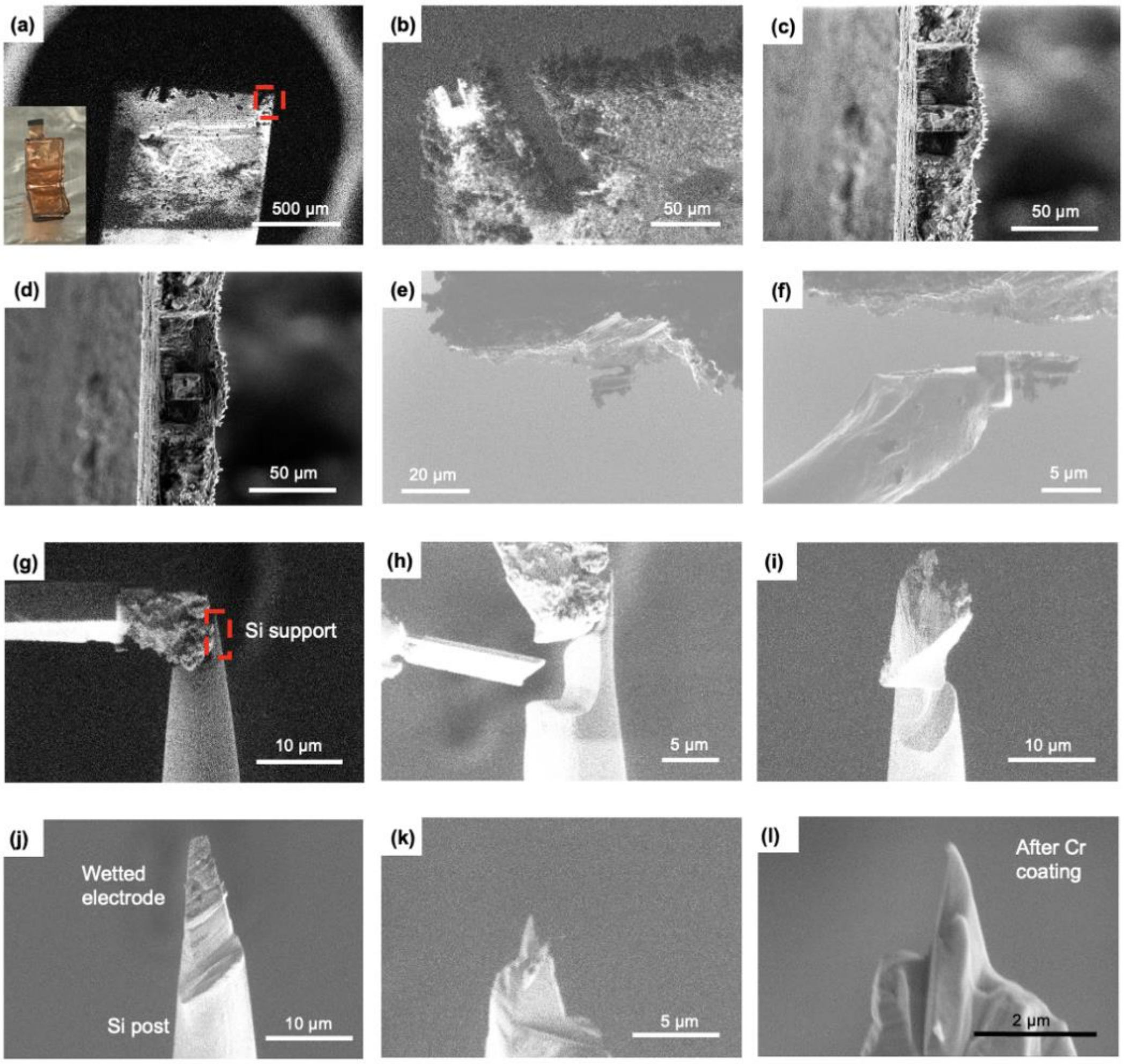

**Figure 5 – Upright mounting and APT specimen preparation from a graphite electrode wetted with uncycled electrolyte.** (a) Upright-mounted electrode with the edge ROI marked in red and the mounting configuration shown in the inset, (b) isolation of the ROI through preparation of a finger, (c) – (d) preparation of the cantilever, (e) undercutting at 0° tilt, (f) attachment to the manipulator and lift-out from the bulk electrode, (g) positioning of the lifted-out cantilever onto the pre-prepared Si post, (h) fill-in using Cr, (i) sample on the post prior to sharpening, (j) tip sharpening at 0° tilt, (k) tip sharpening at 52° tilt prior to Cr coating, (l) final APT tip after coating, with a tip diameter of ~250 nm.

The prepared needle was subsequently analysed by APT, as shown in Fig. 6. The detector event histogram shows a relatively continuous distribution of detected ions, with no pronounced hotspots (Fig. 6(a)), although a gradual variation in detector intensity is observed across the map, which may reflect minor laser-induced asymmetry[52]. The voltage profile, shown in Fig. 6(b), generally increases throughout the experiment, reaching ~5 kV, with several changes associated with adjustments to the acquisition conditions. The corresponding mass spectrum contains a large number of peaks in the range 0 – 80 Da (Fig. 6(c)).

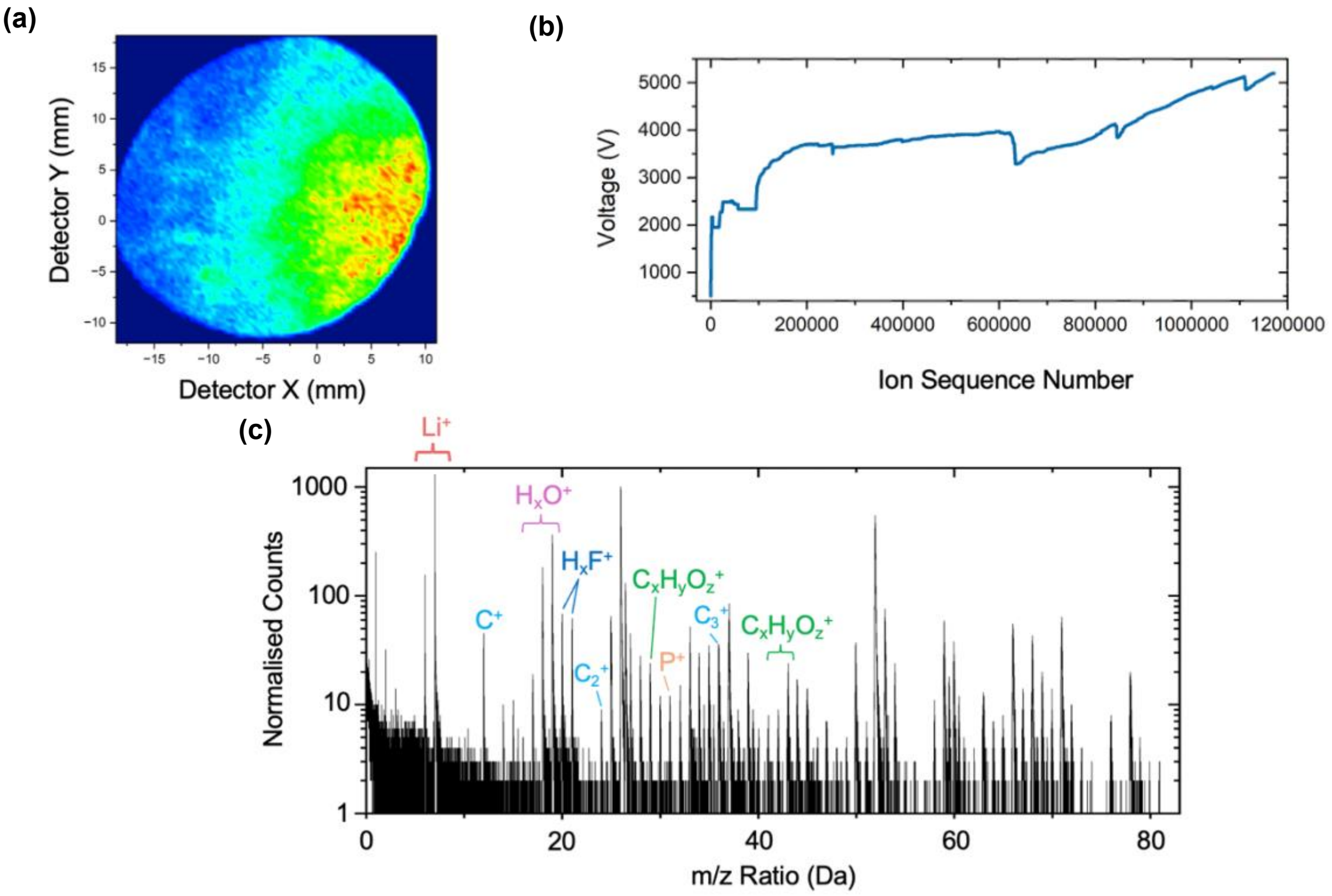

**Figure 6 – APT analysis of the graphite electrode wetted with uncycled electrolyte.** (a) Detector event histogram, (b) voltage profile as a function of ion sequence number, and (c) corresponding mass spectrum over the 0 – 80 Da range with selected peaks assigned, including representative $C_xH_yO_z^+$ species.

The mass spectrum contained $Li^+$ at 6 and 7 Da, together with several species consistent with the presence of the electrolyte used in this study (1.0 M $LiPF_6$ in 1:1 v/v EC:EMC). These included $H_xF$ and P-containing signals at 20 – 21 Da and 31 Da, respectively, as well as $C_xH_yO_z$ molecular species at 29, 32 – 33, 41 – 43, 45 – 47 Da, among other m/z ranges[53]. Further details of the peak ranging are provided in Table S1 of the Supplementary Information. To distinguish peaks originating from the graphite electrode from those in the electrolyte, particularly C-containing species, and to aid interpretation of their clustering behaviour, this dataset was compared with that of an APT specimen prepared from pristine graphite (that had not been exposed to the electrolyte), shown in Fig S6. Alongside several $C_x^+$ peaks, various $C_xH_y$ peaks were also detected in the pristine graphite specimen, likely from adventitious carbon (surface contamination by hydrocarbons originating from the surrounding environment), which is typical of air-transferred samples and has been widely reported[54–56]. Consequently, $C_xH_y^{n+}$ species detected in the uncycled electrolyte dataset were not assigned to the electrolyte, as residual adventitious carbon could also have been present, despite the use of an inert transfer workflow. Overall, the presence of electrolyte-related species provides an initial indication that material from the deliberately wetted region was retained during preparation.

Fig. 7 presents the reconstructed APT dataset from the upright-mounted graphite electrode wetted with uncycled electrolyte. The spatial distributions of the key phases are shown in Fig. 7(a) – (b). The Li contour map in Fig. 7(a) shows a Li-rich region towards the top of the reconstructed volume, above the Cr-rich phase associated with the fill-in used to secure the specimen to the Si post. The corresponding contour maps, Fig. 7(b), show that Li co-localises with $C_xH_yO_z$, $H_xF$ and P species, strongly indicating the presence of an electrolyte phase in the top region of the needle. These electrolyte-associated species are also present within and co-localised with the $C_x$-rich region attributed predominantly to the graphite substrate. The $C_x$-rich phase is positioned adjacent to the Cu-rich phase which corresponds to the current collector.

The co-localisation of Li and $C_xH_yO_z$ is shown more clearly in Fig. 7(c), supporting the assignment of these regions to the electrolyte. Their overlapping distribution also indicates that substantial phase segregation was not introduced during freezing. The presence of electrolyte-associated species within the graphite-rich region is consistent with electrolyte permeation into the porous electrode, while their greater concentration towards the top of the reconstruction indicates that electrolyte was also retained as a discrete deposit at the electrode edge. Importantly, the resulting specimen therefore contains both the edge-localised electrolyte deposit and the underlying multiphase electrode structure within a single APT analysis volume. This demonstrates that the upright-mounting procedure can be used not only to isolate a feature of interest from the electrode edge, but also to preserve its spatial relationship with the underlying electrode during APT specimen preparation. A simplified representation of the reconstructed specimen and the arrangement of the identified phases is shown in Fig. 7(d). The corresponding bulk compositional analysis of this dataset is provided in Fig. S7.

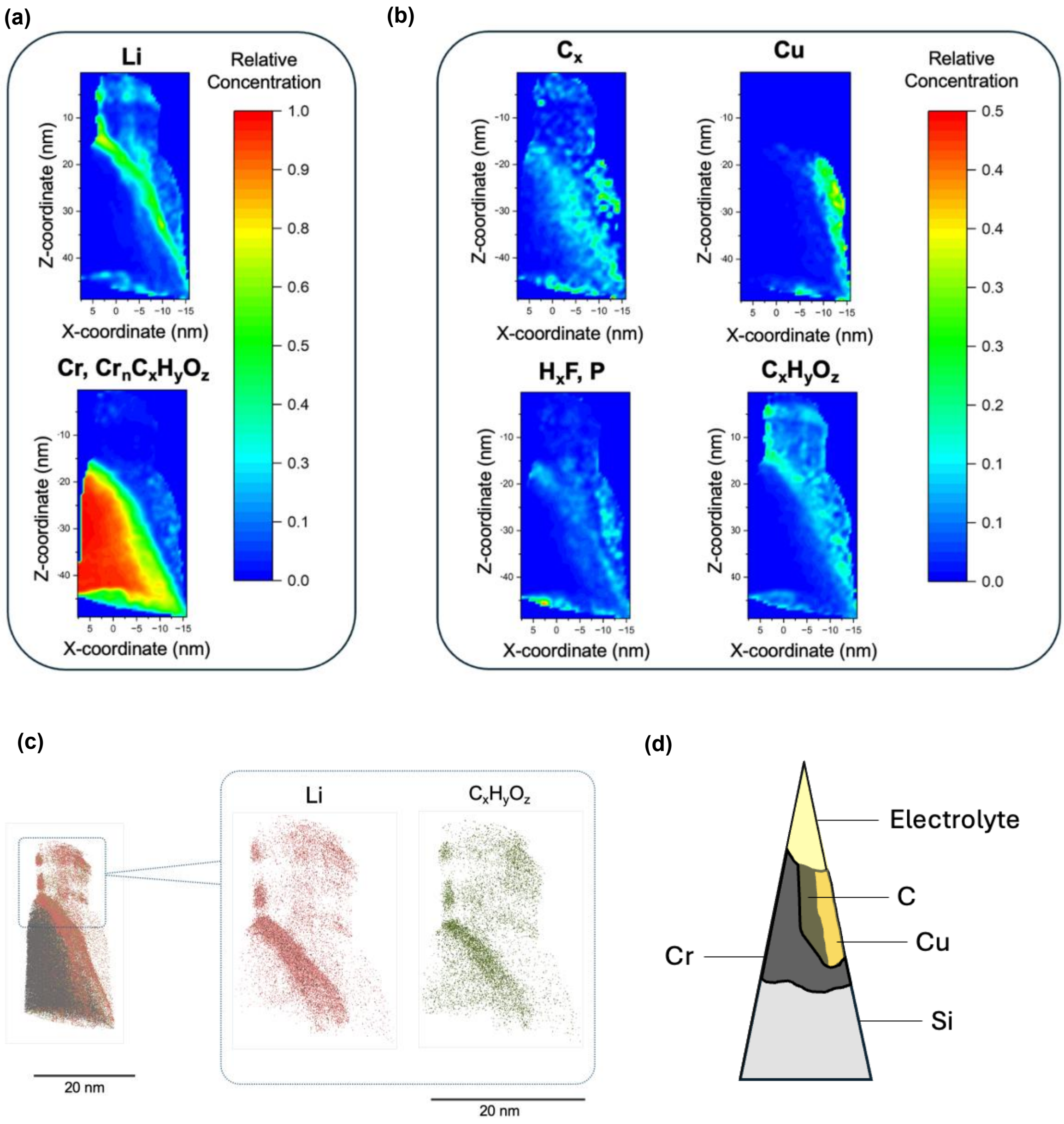

**Figure 7 – Spatial distribution of the key phases in the reconstructed dataset from the graphite electrode wetted with uncycled electrolyte.** (a) Contour maps of Li and Cr-containing species, (b) contour maps of $C_x$, Cu, $H_xF$ and P, and $C_xH_yO_z$ species, (c) 3D ion maps of the ROI enclosed in the blue region to highlight the co-localisation of Li and $C_xH_yO_z$ species, (d) simplified schematic illustrating the spatial arrangement of the notable phases within the analysed volume.

## 3.3 Application to Edge-Localised Li Dendrites

Following validation using the electrolyte-wetted graphite electrode, the upright-mounting procedure was applied to a cycled graphite electrode containing edge-localised Li dendrites (Fig. 8). This provided a more scientifically relevant application of the approach, in which the feature of interest was formed electrochemically during cycling rather than introduced deliberately as part of a model system. The aim was therefore to determine whether the method could be used to site-specifically isolate Li plated along the graphite electrode perimeter and prepare the targeted region for APT analysis.

Formation was performed between 1.0 – 0.005 V vs. $Li^+/Li$, during which using the *operando* ECC-Opto-10 cell. During this process, the graphite is lithiated and delithiated at a low charge rate to facilitate the formation of the SEI on the graphite through reductive decomposition of the electrolyte. Subsequently, the graphite electrode was lithiated at a current density of 0.30 mA $cm^{-2}$ to a cut-off potential of -0.05 V vs. $Li^+/Li$. An optical image of the cycled electrode, showing dendrites concentrated around its perimeter, is presented in Fig. 2(a), with the corresponding voltage profile shown in Fig. S8. This preferential plating along the edge is likely associated with local variations in current distribution arising from the electrode geometry, as reported in previous studies[25,57]. The preparation sequence used to obtain a lift-out from the edge region is shown in Fig. 8(a) – (e). Following cryogenic preservation and upright mounting, the cantilever was prepared at 52° along the edge of the electrode where dendrites were concentrated. Once the sample was lifted out and attached to the Si post, Cr was used to secure this attachment, and successive milling was used to prepare the APT specimen. The final needle is shown in Fig. 8(f) after coating the sample with Cr. Importantly, the upright configuration here provided direct access to the edge where these dendrites were concentrated, allowing the ROI to be isolated without relying on access through the principal electrode surface. Despite successive milling, the final specimen retained some shouldering, highlighting the practical difficulty of preparing heterogeneous, carbon-rich electrodes into the sharp needle geometry required for APT[58].

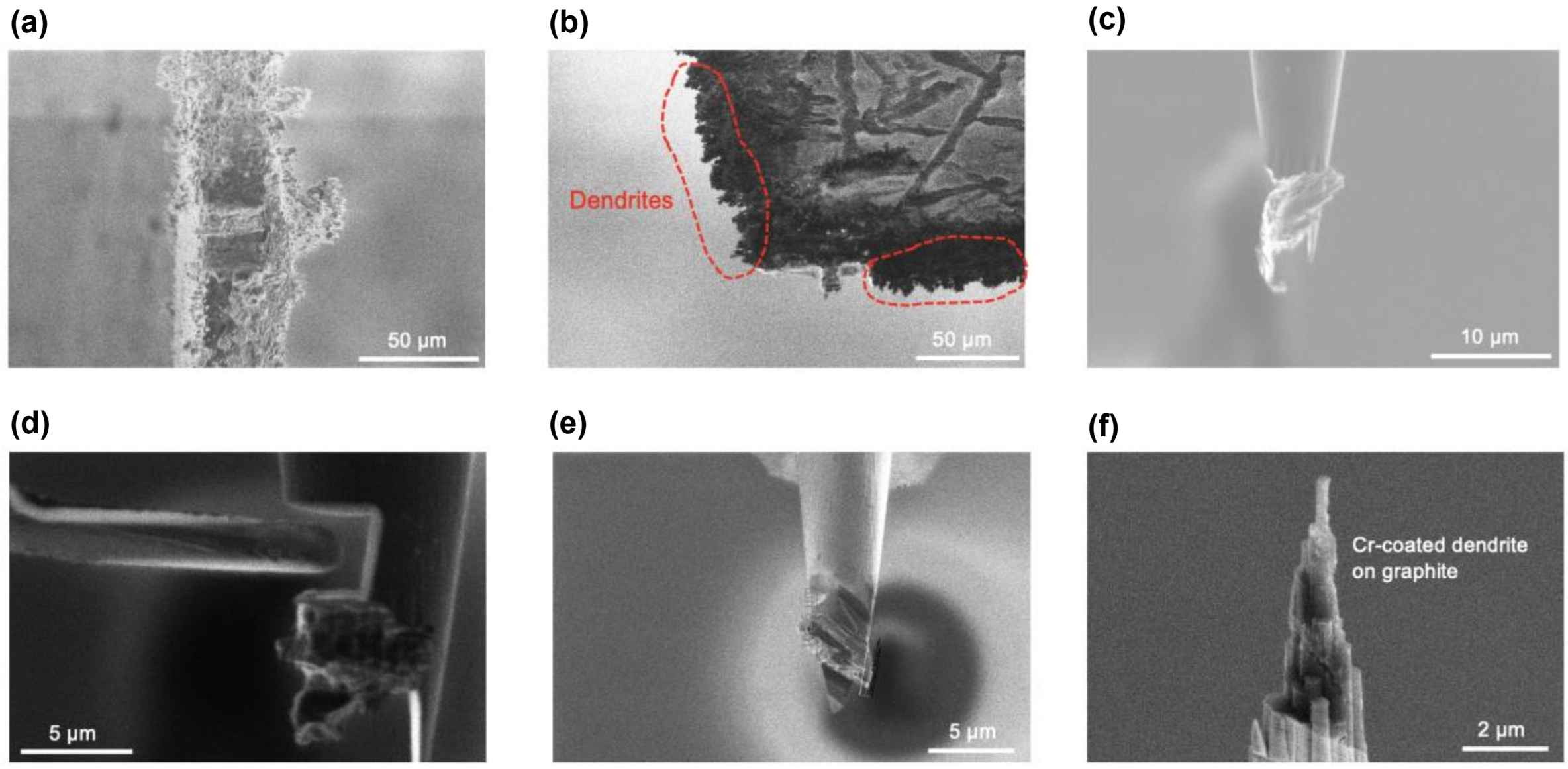

**Figure 8 – Site-specific cryo-FIB preparation of an APT specimen from edge-localised Li dendrites on a cycled graphite electrode.** (a) Upright-mounted electrode edge containing the targeted Li dendrites, (b) preparation of the cantilever from the edge-localised region, where dendrites are circled in red, (c) lifted-out specimen positioned on the pre-prepared Si post, (d) Cr fill-in used to secure the specimen to the post, (e) specimen following Cr fill-in and initial shaping, and (f) final Cr-coated needle.

Fig. 9 presents the APT analysis of this specimen. The mass spectrum, shown in Fig. 9(a), is dominated by $Li^+$ at 6 – 7 Da, alongside Cr-containing species at 26 and 52 Da. Unlike the electrolyte-wetted graphite system, no electrolyte-associated species, including P, $H_xF$, and $C_xH_yO_z$, were detected. Combined with the site-specific preparation of the plated electrode edge, these results are consistent with the successful isolation of a Li dendrite (noting that any surrounding SEI/ electrolyte species may lie outside of the analysed volume). The corresponding 3D ion map and contour maps, shown in Fig. 9(b) – (c), show the spatial distribution of the Li- and Cr-rich phases. The reconstruction was generated using shank-angle mode rather than voltage mode, strictly for visualisation purposes. The Li-rich region is clearly positioned above the Cr-rich fill-in, highlighting the retention of the targeted region during specimen preparation. Together with the validation using the electrolyte-wetted graphite test system, these results demonstrate the feasibility of the upright-mounting approach for site-specific cryogenic APT preparation of edge-localised features. More broadly, the procedure provides a simple and adaptable route for extending site-specific APT preparation for edge regions across a vast range of material systems.

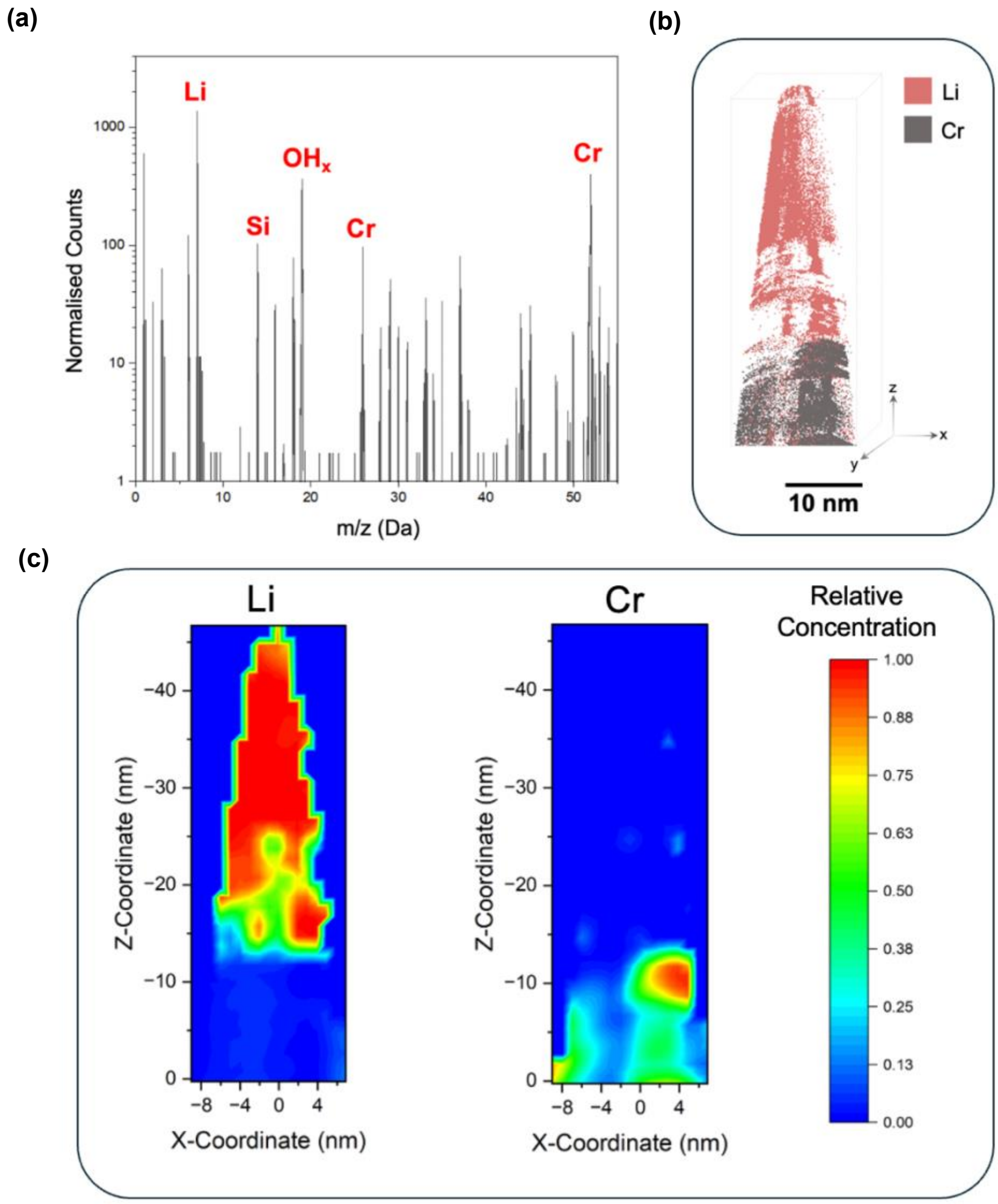

**Figure 9 – APT analysis of the specimen prepared from the edge-localised Li dendrites.** (a) Mass spectrum over the 0 – 55 Da range, showing prominent Li, $OH_x$, Cr, and Si species. (b) 3D ion map showing the spatial distribution of Li and Cr species, and (c) corresponding concentration contour maps of Li and Cr.

# 4 CONCLUSION

This work demonstrates a site-specific cryogenic specimen preparation workflow for atom probe tomography (APT), enabling the analysis of features localised along the edges of LIB electrodes. The approach uses a simple upright-mounting procedure, in which folded Cu tape secures the electrode perpendicular to a conventional cryogenic sample holder, providing direct access to regions that are otherwise difficult to target using conventional flat mounting. Importantly, this geometry also allows different phases to be positioned sequentially along the APT field-evaporation axis, rather than laterally across the specimen. This approach is particularly advantageous under cryogenic conditions, where conventional GIS deposition is more challenging and the limited rotational capabilities of some cryogenic micromanipulators can restrict specimen manipulation during FIB preparation. The procedure requires no specialised mounting hardware or modifications to the FIB stage and maintains mechanical stability throughout repeated cryogenic transfers.

We first demonstrated the approach using a pristine graphite electrode wetted with uncycled electrolyte. Site-specific cryogenic PFIB preparation enabled an edge-localised electrolyte deposit and the underlying graphite and Cu current collector to be incorporated into a single APT specimen. Subsequent analysis identified electrolyte-associated species, including Li, $H_xF$, P, and $C_xH_yO_z$, demonstrating that the workflow can retain and analyse chemically distinct phases within the same specimen.

The method was then applied to an over-lithiated graphite electrode containing Li dendrites concentrated around its perimeter. Using the upright configuration, we were able to target and prepare a Li-rich region for cryogenic APT analysis. The resulting mass spectrum and three-dimensional reconstruction were consistent with the presence of targeted Li dendrites, demonstrating that the procedure can also be applied to localised features formed during electrochemical cycling.

Overall, this work provides a straightforward extension to existing cryogenic APT workflows, allowing regions that are difficult to access using conventional mounting geometries to be targeted and analysed. Although demonstrated here using lithium-ion battery electrodes, the approach is not limited to electrochemical systems and could be applied to other air- and beam-sensitive materials containing localised deposits or interfaces[59–61]. More broadly, it highlights the importance of specimen mounting and orientation in determining which regions of a material can be accessed and subsequently analysed by APT.

# AUTHOR CONTRIBUTION

Syeda Ramin Jannat: Conceptualisation; methodology; investigation; formal analysis; data curation; visualisation; validation; writing—original draft. James O. Douglas: Methodology; investigation; writing—review and editing. Neil Mulcahy: Methodology; investigation; writing—review and editing. Lukas Worch: Methodology; investigation; writing—review and editing. Aigerim Omirkhan: Methodology; investigation; writing—review and editing. Zhenyu Guo: Methodology; investigation; writing—review and editing. Yuanzhu Zhao: Methodology; investigation; writing—review and editing. Maria-Magdalena Titirici: Resources; writing—review and editing. Ifan E. L. Stephens: Resources; supervision; writing—review and editing. Mary P. Ryan: Supervision; funding acquisition; writing—review and editing. Baptiste Gault: Conceptualisation; methodology; supervision; writing—review and editing. Michele Shelly Conroy: Conceptualisation; methodology; investigation; supervision; funding acquisition; project administration; writing—review and editing.

## ACKNOWLEDGEMENTS

This research was supported by the Facility for Cryo-Enabled Multi-microscopy for Nanoscale Analysis in the Engineering and Physical Sciences (Cryo-EPS), EP/V007661/1. S.R.J. and M.P.R. gratefully acknowledge funding from UKRI under the Engineering and Physical Sciences Research Council (EPSRC) Faraday Training Grant, EP/S514901/1. N.M., A.O., M.P.R., and M.S.C. acknowledge funding from the EPSRC and Shell through the InFUSE Prosperity Partnership (EP/V038044/1). L.W. thanks the EPSRC Centre for Doctoral Training in the Advanced Characterisation of Materials (CDT-ACM, EP/S023259/1) for their PhD studentship funding. I.E.L.S. acknowledges the Faraday Institution (grant FIRG025). M.P.R. also acknowledges support from the Armourers and Brasiers Company. M.S.C. is thankful for support and funding from the Royal Society Tata University Research Fellowship (URF\R1\201318), Royal Society Enhancement Award (RF\ERE\210200EM1), and ERC CoG DISCO grant 101171966.

## CONFLICT OF INTEREST STATEMENT

All authors declare no known conflicts of interest.

## DATA AVAILABILITY STATEMENT

The data presented in this study are included in the main article and Supplementary Information.

# Supplementary Information

## ECC-Opto-10 Cell Assembly and Set-Up

To prepare the cycled graphite electrode used in this study, an ECC-Opto-10 *operando* cell (EL-Cell) was used, configured in the 'side-by-side' arrangement. In other words, the NMC811 and graphite electrodes were positioned laterally adjacent to each other with a 10 mm-diameter GF/A separator placed underneath (thickness of 0.26 mm), supplied by Merck KGaA (Darmstadt, Germany). A 4 mm-diameter Li reference electrode was positioned beneath this separator on top of the Li reference pin. Subsequently, 100 µL of electrolyte (1.0 M $LiPF_6$ in 1:1 v/v EC:EMC), also supplied by Merck KGaA (Darmstadt, Germany), was pipetted onto the electrodes and separator before the top unit (including the lid, window thrust ring, window flange, PE window seal, and window) was screwed in place on top of this. Note that a torque screwdriver, set to 0.2 Nm, was used to attach the window flange to the lid to ensure reproducible assembly. The sapphire window (thickness of 0.3 mm, diameter of 22 mm, respectively) sat directly on top of the electrodes in the cell.

Fig. S1 shows the experimental set-up used for *operando* visualisation, comprising the cell itself, the optical microscope, monitor, and potentiostat.

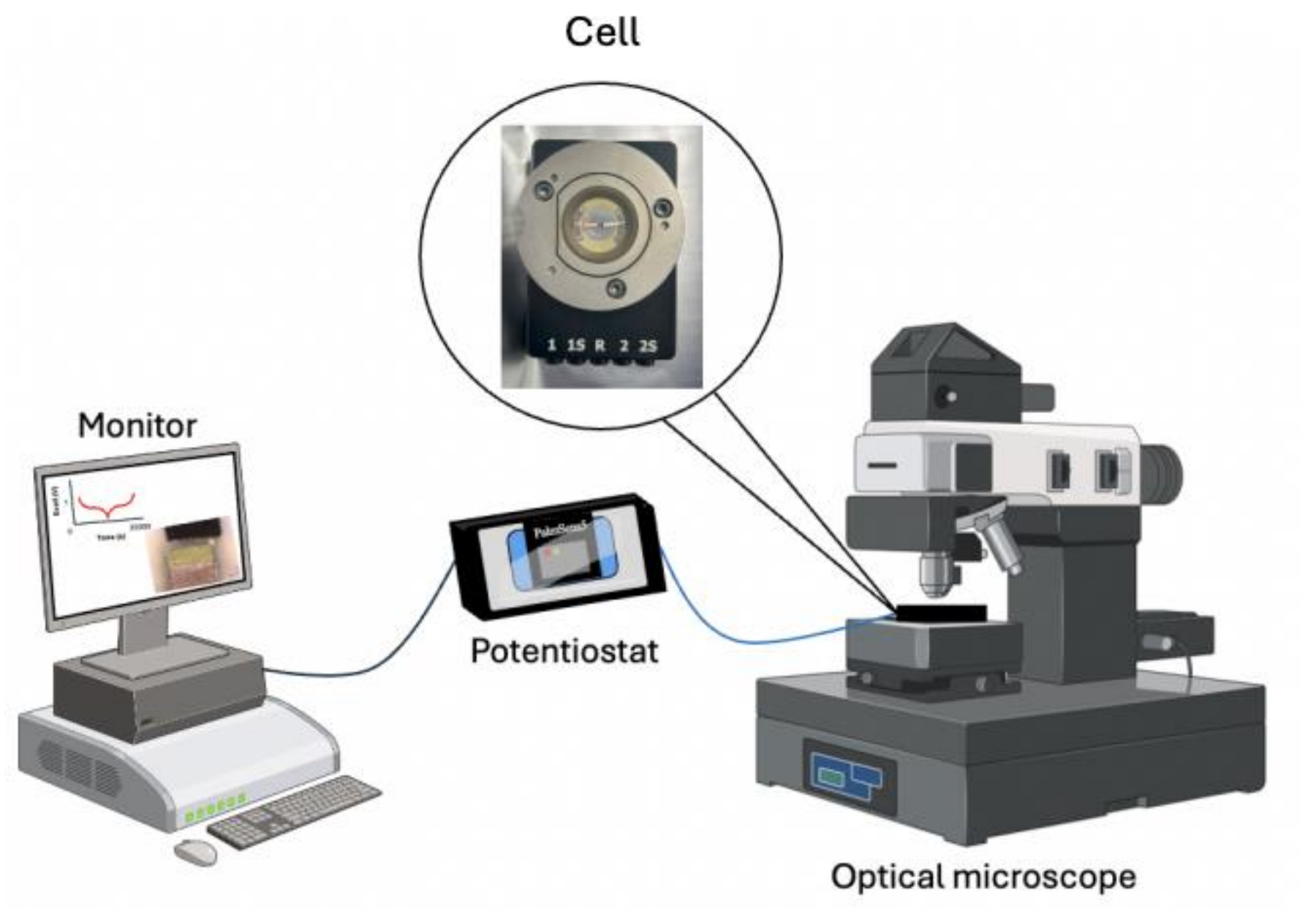

**Figure S1 -** Experimental set-up used for *operando* visualisation of the electrochemical cell, showing an assembled cell, the optical microscope, monitor and potentiostat. Schematic partially created in BioRender. Jannat, S.R. (2026) https://BioRender.com/hv8xarn.

## Cryogenic Workflow

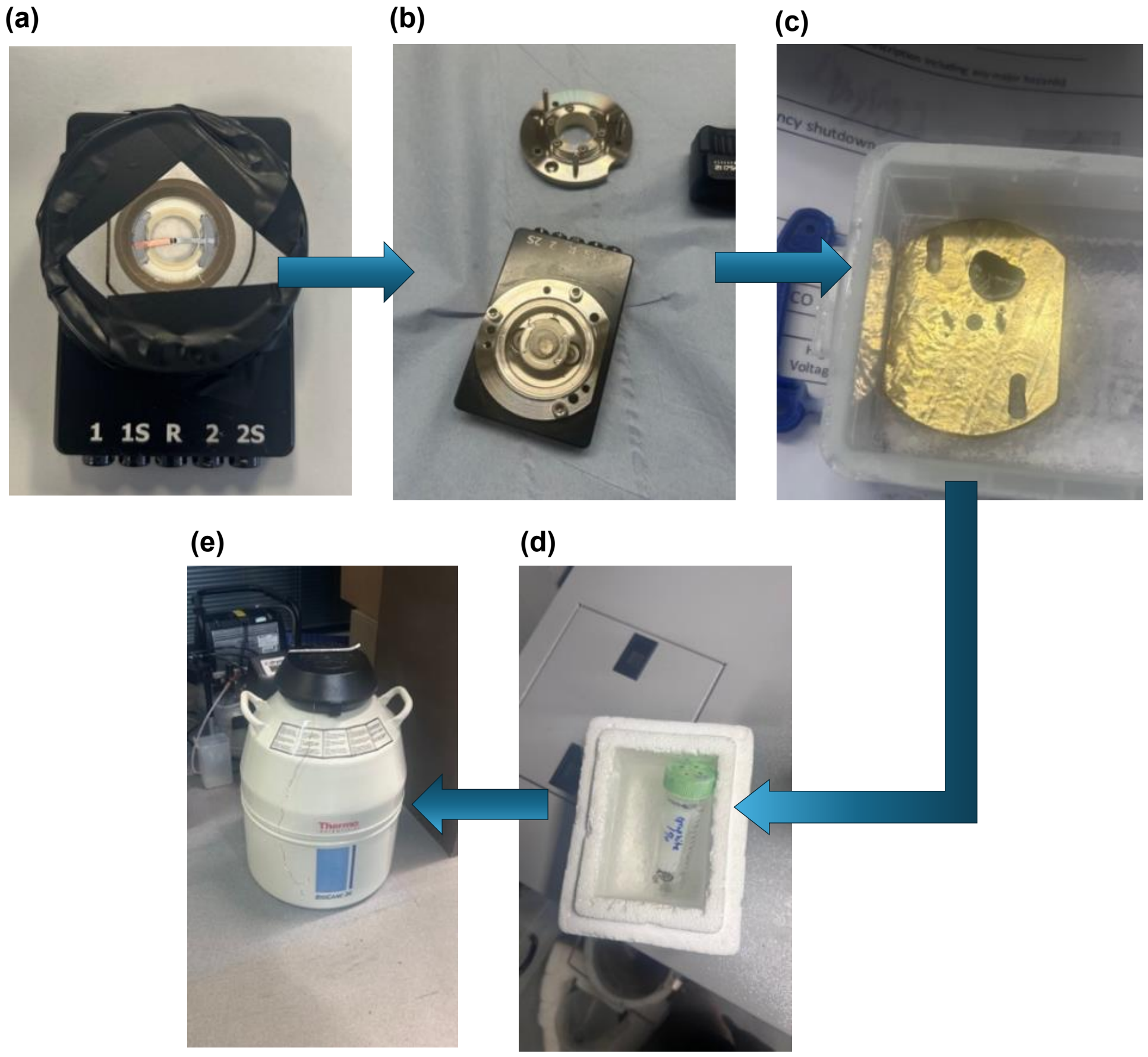

**Figure S2 – Sample retrieval and storage following electrochemical cycling.** (a) The *operando* ECC-Opto-10 cell used for electrochemical cycling, (b) the disassembled cell following cycling, from which the electrode of interest was carefully extracted, (c) controlled cooling of a representative electrode (shown for illustration) on an $LN_2$-pre-cooled block, (d) transfer of the frozen electrode into a vented centrifuge tube, and (e) storage under $LN_2$ in a cryogenic dewar.

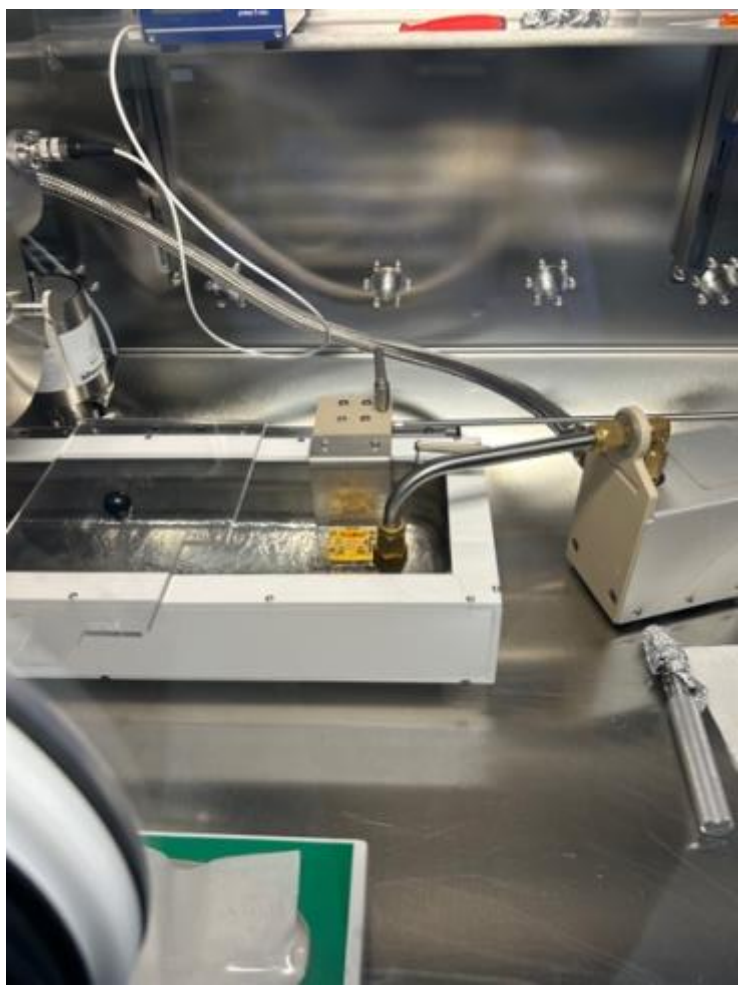

**Figure S3 –** Stainless steel basin within the $N_2$-glovebox during filling with $LN_2$ from a pressurised dewar.

**(a)** **(b)**

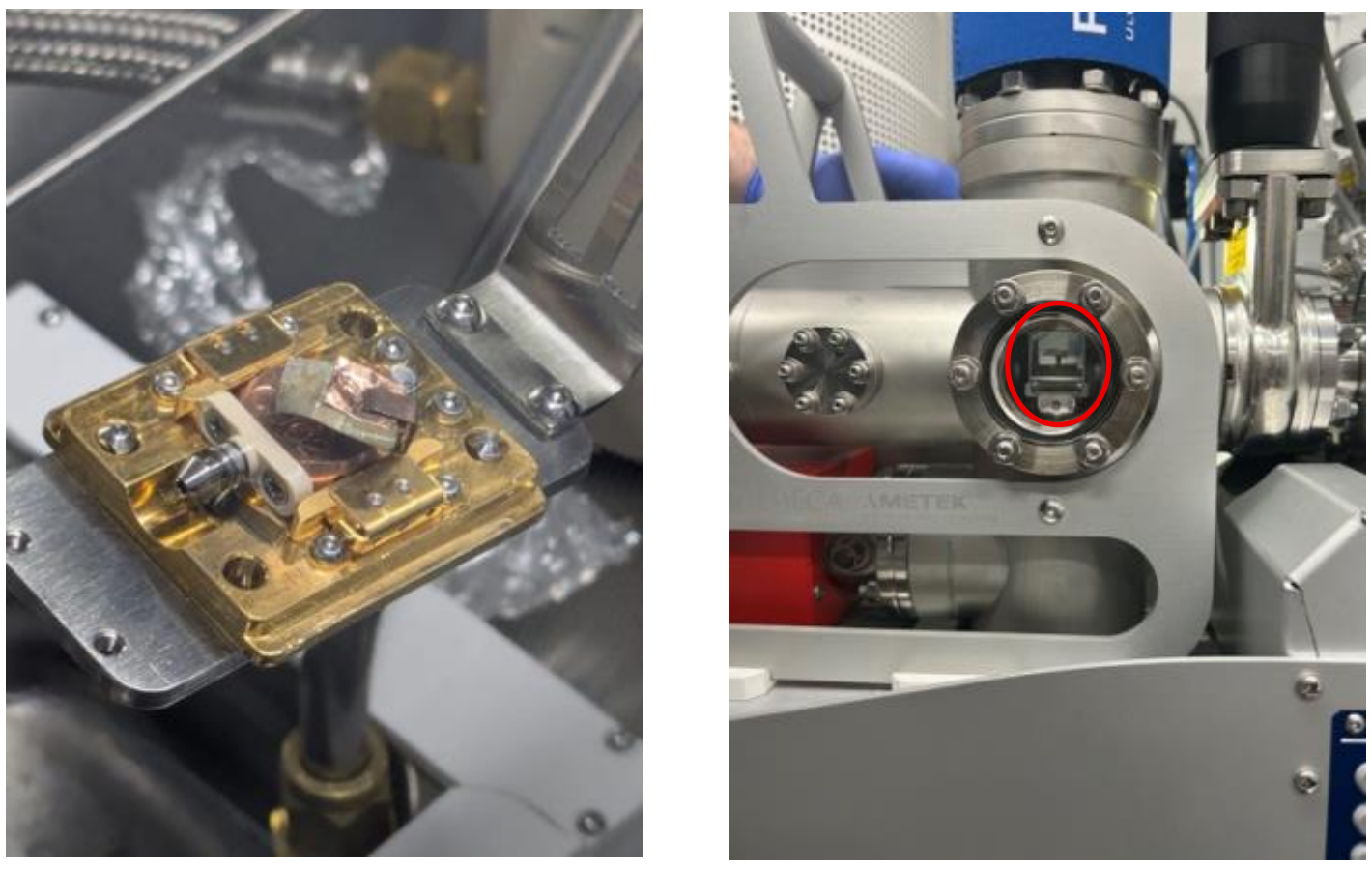

**Figure S4 – Cryogenic sample transfer from the glovebox.** (a) Cryo-puck positioned on the transfer stage prior to loading into the load-lock, (b) cryo-puck inside the pre-cooled VCTM, circled in red.

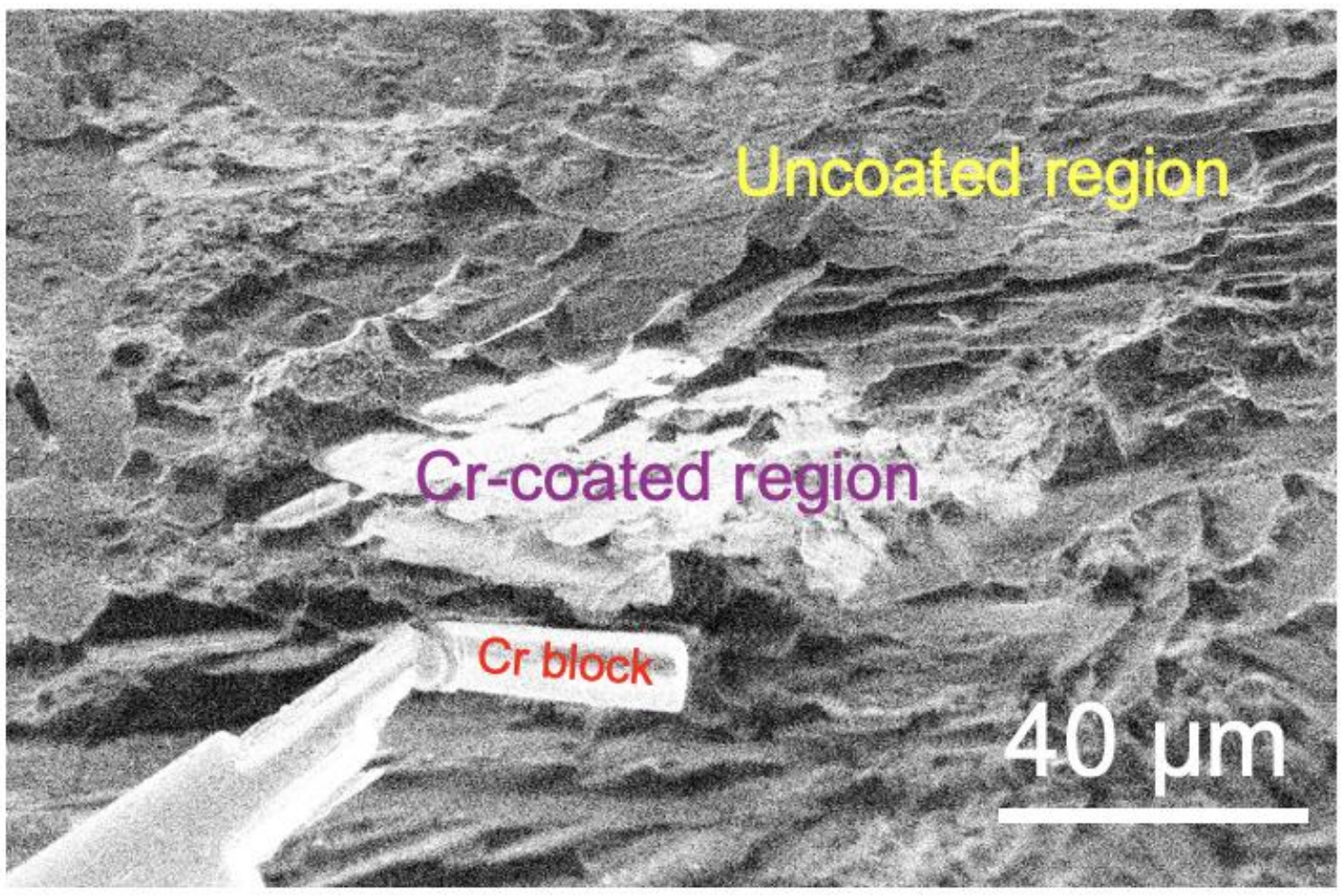

**Figure S5 –** SEM image showing the Cr-coated and uncoated regions of the graphite electrode, together with the Cr block pre-attached to the manipulator using the GIS at room temperature.

## APT of Pristine Graphite Sample

A pristine graphite specimen was prepared as a reference for distinguishing signals originating from the graphite electrode from those associated with the electrolyte in the wetted-electrolyte dataset. To prepare this, a small square of graphite electrode (~0.5 cm x 0.5 cm) was cut out and adhered to an SEM stub (6 mm pin height, Merck KGaA, Darmstadt, Germany) using conductive carbon tape (Agar Scientific Ltd., Rotherham, UK). The sample was prepared under conventional, non-cryogenic conditions, without inert-atmosphere handling. A region of the graphite was simply lifted out using standard FIB procedures, mounted onto a Si post, and sharpened to produce an APT specimen.

The resulting reconstruction and mass spectrum are shown in Fig. S6. The dataset was dominated by $C_x^+$ ($x \leq 4$), which accounted for 84.30 at.% of the detected composition, together with 15.06 at.% O, and 0.68 at.% N, while no Li was detected (consistent with the 'pristine' nature of the sample). The most prominent carbon peak occurred at 12 Da ($C^+$), with smaller $C_2^+$, $C_3^+$, and $C_4^+$ peaks also observed. The negligible intensity above $C_4^+$ indicates the preferential C cluster pathways during field evaporation. Quantitative assignment of C clusters can be notoriously challenging, partly due to overlaps between atomic and molecular ions, e.g. a peak at 24 Da could be $C_2^+$ or $C_4^{2+}$. Therefore, individual carbon species are not deconvoluted and are instead collectively referred to as $C_x$. Several $C_xH_y$ peaks were also present at 26 – 27, 29, and 39 – 41 Da, which are likely associated with adventitious hydrocarbon contamination following preparation and handling under ambient conditions.

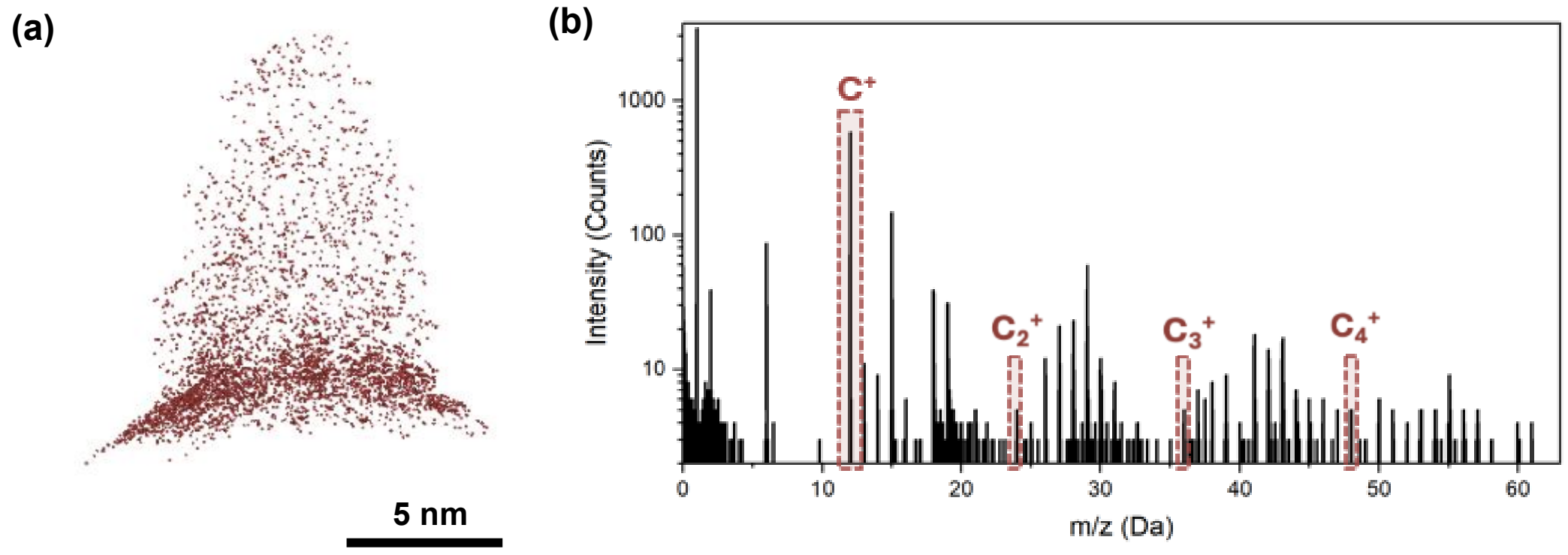

**Figure S6 – APT analysis of the pristine graphite reference dataset.** (a) 3D compositional ion map for $C_x^+$ (x ≤ 4) with a 5-nm scale bar, (b) corresponding mass spectrum with the notable $C_x^+$ peaks indicated.

## APT of the Graphite Electrode Wetted with Uncycled Electrolyte

Table S1 – Notable peak assignments from the uncycled electrolyte dataset.

| Peak Assignment | m/z |
|---|---|
| $Li^+$ | 6 – 7 |
| $C_x^+$ | 12, 24, 36, 48 |
| $H_xO^+$ | 17 – 19 |
| $H_xF^+$ | 20 – 21 |
| $P^+$ | 31 |
| $C_xH_yO_z^+$ | 29, 32 – 33, 41 – 43, 45 – 47, 85 – 88 |

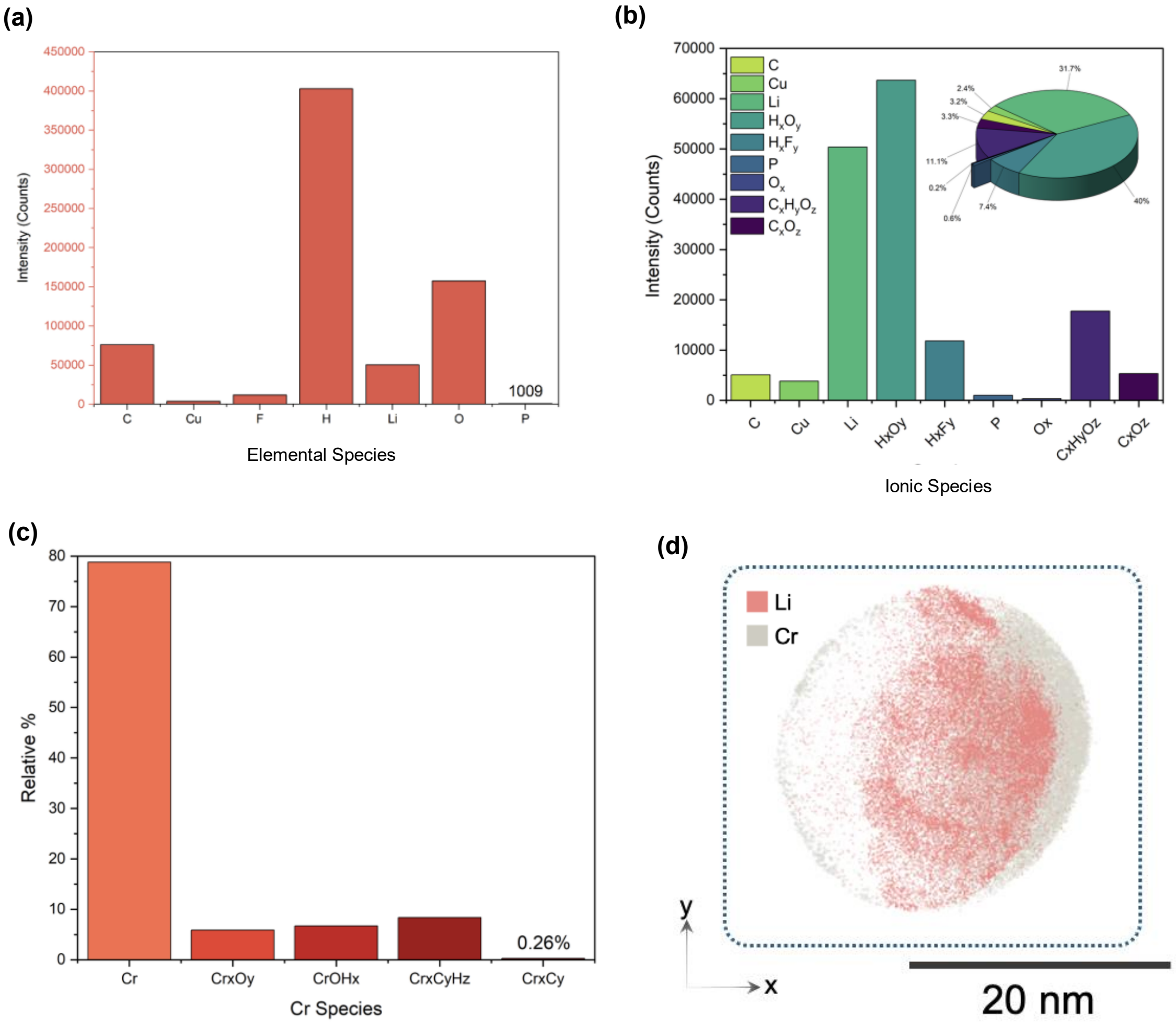

**Figure S7 – Bulk compositional analysis of the APT dataset from the graphite electrode wetted with uncycled electrolyte.** (a) Elemental species distribution, (b) distribution of notable ionic species, with the inset showing their relative contributions to the assigned species count, (c) relative distribution of Cr-containing species, and (d) X-Y ion map showing the spatial distributions of Li and Cr species. The concentrated Cr region on the right-hand side here corresponds to the fill-in to attach the sample to the Si post.

# Electrochemical results for the cycled graphite electrode

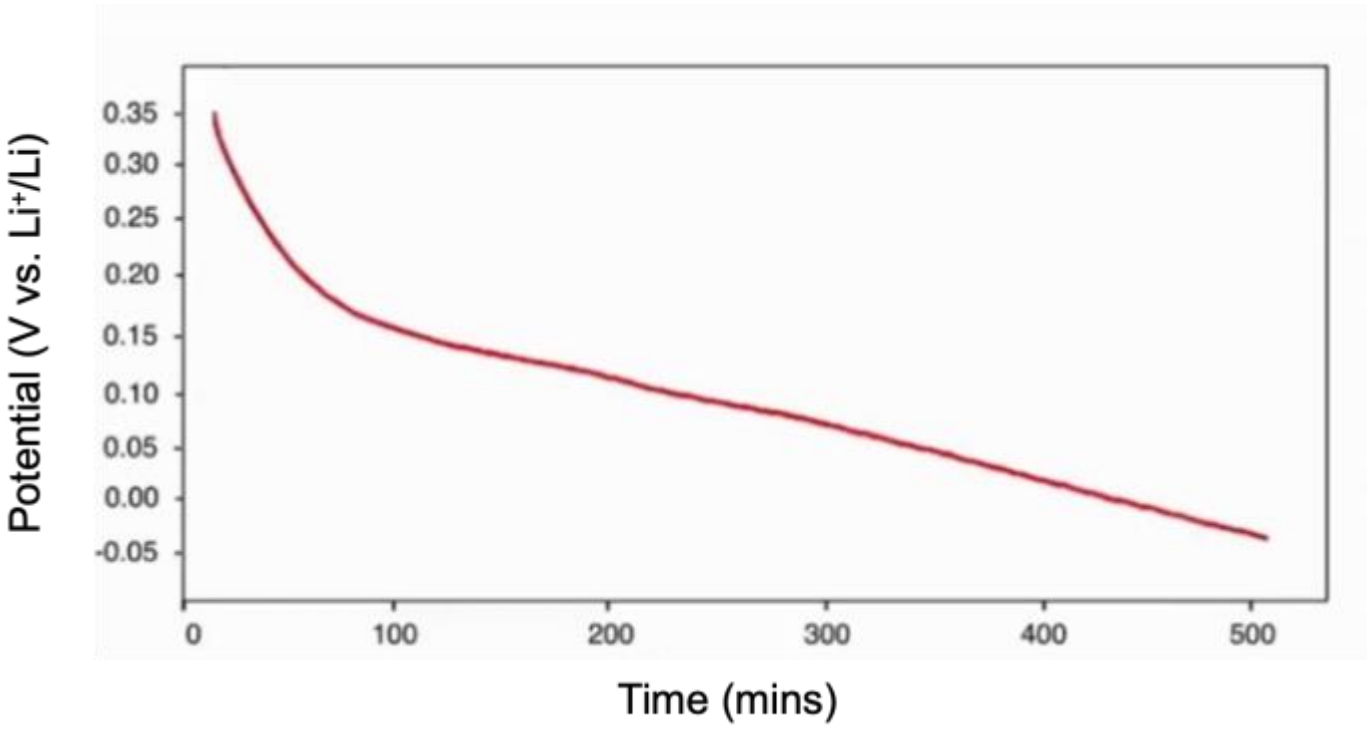

**Figure S8 –** Voltage profile during lithiation of the graphite electrode to a cut-off potential of -0.05 V vs. $Li^+/Li$ at a current density of 0.30 mA $cm^{-2}$ (approximately C/8) using the ECC-Opto-10 *operando* cell set-up at room temperature. The cell comprised graphite as the working electrode, NMC811 as the counter electrode, and Li as the reference electrode, with 1.0 M $LiPF_6$ in EC:EMC (1:1 v/v) as the electrolyte.